\documentstyle [eqsecnum,preprint,aps,epsf,fixes]{revtex}

\def\eps{\epsilon}
\def\svev{\langle s \rangle}
\def\tvev{\langle t \rangle}
\def\stvev{\langle st \rangle}
\def\betat{\beta_{\rm t}}
\def\mod{{\rm mod}}
\def\edge{{\rm edge}}
\def\diag{{\rm diag}}
\def\taud{\tau_{\rm d}}
\def\Nbin{N_{\rm bin}}
\def\bin{{\rm bin}}

\begin {document}

\makeatletter           
\@floatstrue
\def\figure{\let\@capwidth\columnwidth\@float{figure}}
\let\endfigure\end@float
\@namedef{figure*}{\let\@capwidth\textwidth\@dblfloat{figure}}
\@namedef{endfigure*}{\end@dblfloat}
\makeatother

\preprint {UW/PT-96-26}

\title  {Monte Carlo study of very weak first-order transitions in the
   three-dimensional Ashkin-Teller model}

\author {Peter Arnold and Yan Zhang}

\address
    {%
    Department of Physics,
    University of Washington,
    Seattle, Washington 98195
    }%
\date {October, 1996}
\maketitle

\begin {abstract}%
{%
We propose numerical simulations of the Ashkin-Teller model as a foil
for theoretical techniques for studying very weakly first-order phase
transitions in three dimensions.  The Ashkin-Teller model is a simple
two-spin model whose parameters can be adjusted so that it has an arbitrarily
weakly first-order phase transition.  In this limit, there are
quantities characterizing the first-order transition which are universal:
we measure the relative
discontinuity of the specific heat, the correlation
length, and the susceptibility across the transition by Monte Carlo simulation.
\ifpreprintsty
\thispagestyle {empty}
\newpage
\thispagestyle {empty}
\vbox to \vsize
    {%
    \vfill \baselineskip .28cm \par \font\tinyrm=cmr7 \tinyrm \noindent
    \narrower
    This report was prepared as an account of work sponsored by the
    United States Government.
    Neither the United States nor the United States Department of Energy,
    nor any of their employees, nor any of their contractors,
    subcontractors, or their employees, makes any warranty,
    express or implied, or assumes any legal liability or
    responsibility for the product or process disclosed,
    or represents that its use would not infringe privately-owned rights.
    By acceptance of this article, the publisher and/or recipient
    acknowledges the U.S.~Government's right to retain a non-exclusive,
    royalty-free license in and to any copyright covering this paper.%
    }%
\fi
}%
\end {abstract}

\section {Introduction}

   One of the modern scenarios for explaining the baryon asymmetry of
the universe depends on the nature of the electroweak phase transition
in the early universe.  The scenario requires the transition to be
first-order, but in some cases of interest the transition is sufficiently
weak and near-critical that simple perturbative expansions around mean
field theory are not well-behaved.  The study of this cosmological
application has spawned a general renewal of interest in techniques
for studying critical or near-critical phase transitions: techniques
that have a two decade history in condensed matter physics.
The Ising model is the canonical example of a three-dimensional
system with a second-order phase transition and has been well studied
both numerically and theoretically.  There has not, however, been
a comparable amount of attention paid to any canonical example of
weak, near-critical first-order transitions.  A simple two-spin
generalization of the Ising model that can provide such a canonical
example---a testbed for theoretical techniques for treating very weakly
first-order transitions in three dimensions---is the Ashkin-Teller model.
By tuning parameters in this model, one can obtain first-order transitions
that are arbitrarily weak.

   A full discussion of our motivation for studying this model, and its
similarities and dissimilarities with the electroweak phase transition, is
given in ref.~\cite{summary}.  In the present paper, we will study
the relative discontinuity of various physical quantities across the
first-order transition.  In particular, we measure the ratios
$C_+/C_-$, $\xi_+/\xi_-$, and $\chi_+/\chi_-$
where $C_+$, $\xi_+$, $\chi_+$ are the
specific heat, correlation length, and susceptibility in the disordered
(high-temperature) phase and $C_-$, $\xi_-$, and $\chi_-$ are the same in the
ordered (low-temperature) phase.  These ratios are universal in the limit
that the first-order transition is arbitrarily weak and provide a set of
tests against which to measure theoretical techniques for studying
weakly first-order transitions.
In particular, a comparison of our numerical results against the predictions of
$\eps$-expansion methods \cite{eps,rudnick} may be found in
ref.~\cite{summary}.

In the remainder of this introduction, we review the Ashkin-Teller model
and then present our final results.  We will review the parameter $x$ of
the Ashkin-Teller model whose $x{\to}0^+$ limit yields arbitrarily
weakly first-order transitions.
In sec.~\ref{sec:method}, we give a broad overview of our
method for making measurements in the two phases and for
determining the transition temperature.
In sec.~\ref{sec:analysis}, we present our measured ratios for
an assortment of values of $x$; discuss how $C_\pm$, $\xi_\pm$,
and $\chi_\pm$ should scale at small $x$; explain how finite-$x$ corrections
to our $x{\to}0^+$ ratios should scale with $x$; and discuss our
procedure for extracting the $x{\to}0^+$ limit.
In sec,~\ref{sec:details}, we discuss the details that went into
the individual measurements at each $x$ such as our procedure for
measuring susceptibilities and correlation lengths, our assessment
of finite volume errors, and
uncertainties in the transition temperature and their effect
on our measurements.


\subsection {Review of the Ashkin-Teller Model}

\begin {figure}
\vbox
    {%
    \vspace* {-17pt}
    \begin {center}
        \leavevmode
        
        \epsfbox [150 50 500 800] {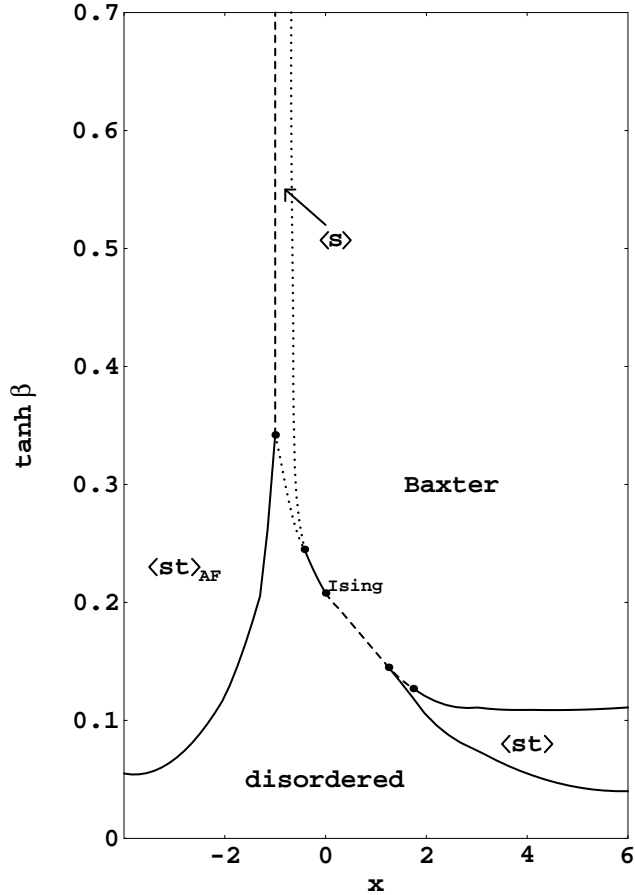}
    \end {center}
    \caption
        {%
        \label {fig:phase-diagram}
        Believed phase diagram of three-dimensional Ashkin-Teller
        model on a simple
        cubic lattice, taken from
        ref.~\protect\cite{ditzian}, where it was extracted from series
        analysis and Monte Carlo data.  Dashed and solid lines indicate
        first and second-order transitions respectively.  Dotted lines
        indicate cases where the nature of the transition has not been
        unambiguously determined.  The phases are
        labeled disordered ($\svev=\tvev=0$); Baxter (ferromagnetic with
        $\svev$, $\tvev$, and $\stvev$ non-zero); ``$\stvev$'' (where
        $\stvev$ is ferromagnetically ordered but $\svev=\tvev=0$);
        ``$\stvev_{\rm AF}$'' (the same but anti-ferromagnetically
        ordered); and ``$\svev$'' (where either $\svev$ or $\tvev$ is
        ferromagnetically ordered but the other is not and
        $\stvev=0$.)
        }%
    }%
\end {figure}

   The (symmetric) Ashkin-Teller model \cite{ashkin&teller}
is a system with two Ising spins $s_i$ and $t_i$
per lattice site $i$, with the nearest-neighbor interaction
\begin {equation}
   \beta H = - \beta \sum_{\langle ij \rangle}
      (s_i s_j + t_i t_j + x s_i t_i s_j t_j) \,,
   \qquad\qquad
      s_i, t_j = {\pm 1} \,.
\end {equation}
Special cases of interest include $x{=}0$, which corresponds to two
decoupled Ising models, and $x{=}1$, which can be rewritten as
\begin {equation}
   \beta H(x{=}1) = - 4\beta \sum_{\langle ij \rangle}
      \left(\delta_{s_is_j}\delta_{t_it_j} - {\scriptstyle{1\over4}} \right)
\label{eq:potts-form}
\end {equation}
and is equivalent to the 4-state Potts model \cite{potts}.
The phase diagram of the model in three dimensions on a simple cubic
lattice is sketched in
fig.~\ref{fig:phase-diagram} \cite{ditzian}.
The portion of this diagram we will focus on
is the neighborhood of the Ising tri-critical point at $x{=}0$.
By studying arbitrarily small but positive values of $x$, we can study
arbitrarily weak first-order phase transitions.  Our goal will be to
extract ratios such as
\begin {equation}
   \lim_{x\to0^+} \>
   {C_+\over C_-} \equiv
   \lim_{x\to0^+} \> {\lim_{\beta\to\betat^-} \> C \over
                      \lim_{\beta\to\betat^+} \> C} \,,
\label{eq:ratio def}
\end {equation}
where $\betat=\betat(x)$ is the inverse transition temperature for a given
value of $x$.

   For generic values of the parameter $x$, the internal symmetry group of the
model is that of a square, $D_4$, acting on the spin states $(s,t) = (++),
(+-), (--), (-+)$.  The line $x{=}0$ is a line of enhanced symmetry: the
translation symmetry decouples for the $s$ and $t$ spins.%
\footnote{
   This makes the Ising point a particularly attractive multi-critical
   point to study because we know it's value of $x$ in advance.  Other
   multi-critical points would first require a numerical search for the
   corresponding multi-critical value $x_{\rm c}$ of $x$ before we could
   proceed to numerically extract the limit $x{\to}x_{\rm c}$ of
   ratios analogous to (\ref{eq:ratio def}).
}
(A simple continuum model with the same long-distance
degrees of freedom and symmetries is the cubic anisotropy model,
a two-scalar field theory discussed in
refs.~\cite{summary,eps,rudnick,misc-cubic,alford}.)


\subsection {Final results}

Our final results for the universal ratios are%
\footnote{
   Due to an improvement in our method of error analysis prior to
   publication, the numbers given here are slightly, but not significantly,
   different than those originally reported in ref.~\cite{summary}.     
}
\begin {eqnarray}
   \lim_{x\to0^+} \> {C_+\over C_-} &=& 0.071(8) \,,
\label{eq:cratio_result}
\\
   \lim_{x\to0^+} \> {\xi_+\over \xi_-} &=& 1.6(2) \,,
\label{eq:xiratio_result}
\\
   \lim_{x\to0^+} \> {\chi_+\over \chi_-} &=& 4.0(6) \,,
\label{eq:chiratio_result}
\end {eqnarray}
where the error estimate on the last result should be taken
with a grain of salt (see sec.~\ref{sec:limit extraction} for details).
The susceptibility $\chi$ is
formally defined in the infinite-volume limit by adding a source term
$- \sum_i {\bf h} \cdot {\bf S}_i$
to $\beta H$, where ${\bf S} \equiv (s,t)$, and then taking
\begin {equation}
   \chi = {1\over2}
      \lim_{{\bf h}\to0} \sum_a {d\over dh_a} \langle S_a \rangle \,.
   \label {eq:chi-def}
\end {equation}

As an incidental consequence of our work, we also have what is, to our
knowledge, the best measurement of the transition temperature of the
4-state Potts model ($x{=}1$) in three dimensions:
\begin {equation}
   \betat\hbox{(4-state Potts)} =
            \cases{
               0.157154(4), & simple cubic; \cr
               0.113752(5), & body-centered cubic; \cr
            }
\end {equation}
where $\beta$ is normalized according to (\ref{eq:potts-form}).
For comparison, earlier
values determined by Monte Carlo and series analysis are given in
Table~\ref{tab:potts}.
The series estimate of ref.~\cite{kim&joseph} for body-centered
cubic (BCC) lattices
is off by roughly thrice their estimated error.

\begin{table}
\def\phan{$\phantom{\simeq}\;$}
\begin {center}
\tabcolsep=6pt

\begin {tabular}{|lll|}                                                  \hline
  & \multicolumn{2}{c|}{$\betat(x{=}1)$}                              \\
  &  simple cubic & \multicolumn{1}{c|}{BCC}                          \\ \hline
Monte Carlo, ref.~\cite{ono&ito} &  \phan 0.158        &              \\
Monte Carlo, ref.~\cite{ohta}    &  ${\simeq}$ 0.15694 &              \\
series, ref.~\cite{kim&joseph}   &  \phan 0.162(2)     &  0.1176(13)  \\
series, ref.~\cite{ditzian&kadanoff}
                                 &  \phan 0.161(3)     &              \\ \hline
\end {tabular}
\end {center}
\caption
    {%
    \label {tab:potts}
    Summary of previous Monte Carlo and series analysis results for $\betat$
    in the 4-state Potts model ($x{=}1$) in three dimensions.
    The Monte Carlo references do not quote error estimates.
    Note that our convention (\protect\ref{eq:potts-form}) for $\beta$
    differs from the usual one by a factor of 4.
    }%
\end{table}


\section {Outline of Calculational Method}
\label{sec:method}

  Our lattices are simple cubic with helical boundary conditions
(periodic except for a twist) and range in sizes up to $480\times120^2$.
In principle, it would be nice to verify the universality of
$x{\to}0^+$
ratios such as $C_+/C_-$ by repeating the calculation on another
lattice type, such as body-centered cubic.  However, we have only
made limited exploration of BCC lattices, and our data
for this case is relegated to Appendix~\ref{app:BCC}.

   We interlace two different update algorithms: a simple heatbath algorithm
and a cluster-update algorithm.%
\footnote{
   But some of our data was generated by heatbath only or by cluster only.
}
The cluster-update algorithm is a simple
generalization of the usual cluster-update algorithms for O($n$) systems
\cite{wolff} and is described in Appendix~\ref{app:cluster}.

\begin {figure}
\vbox
    {%
    \vspace* {-17pt}
    \begin {center}
        \leavevmode
        
        \epsfbox {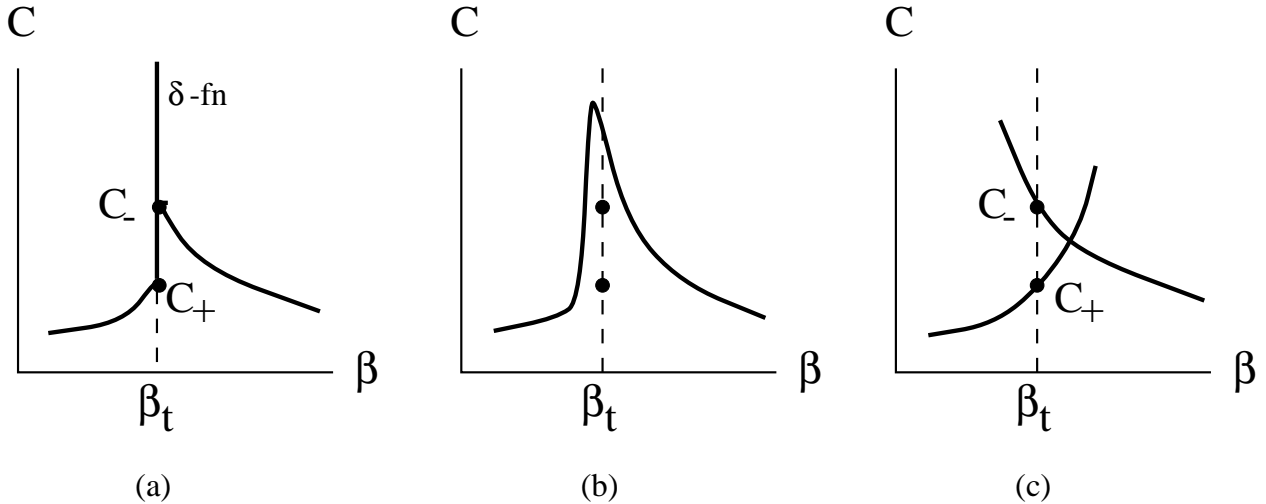}
    \end {center}
    \caption
        {%
        \label {fig:C+C-}
        Specific heat vs.\ $\beta$ at fixed $x$ for (a) infinite volume,
        (b) finite volume, (c) finite-time numerical simulations in
        large volume.  The $\delta$-function in (a) at $\betat$ represents the
        latent heat.  In all cases, $\betat$ represents the true,
        infinite-volume transition temperature.
        }%
    }%
\end {figure}

   Our basic method for extracting our results is best outlined with a
specific example: the specific heat ratio $C_+/C_-$.  The first step is to
measure $C_+/C_-$ at fixed values of $x$, so that we can later extract the
$x{\to}0^+$ limit.  Fig.~\ref{fig:C+C-}a qualitatively sketches the dependence
of specific heat on inverse temperature $\beta$ for fixed $x$.  In finite
volume, the latent heat $\delta$-function at the transition temperature
broadens out into a finite-width peak as in fig.~\ref{fig:C+C-}b.  If we
actually measured this peak in numerical simulations, we would have to attack
the problem of how to extract $C_+$ and $C_-$ from beneath the peak.  However,
the lattices we work on are large enough that the mixing time between the two
phases at the transition is very long compared to the duration of our
simulations.%
\footnote{ This is not an accident.  The cleanest way to eliminate systematic
   errors from finite-volume effects is to work in volumes $V$ large compared
   to the correlation volume $V_\xi$.  But, at the transition temperature, the
   mixing time grows exponentially with $V/V_\xi$.
}
Our results therefore have the form of fig.~\ref{fig:C+C-}c: there is a range
of temperatures, around the transition temperature, in which our simulation
sits in either the ordered or disordered phase depending on whether we start
with ordered or disordered initial conditions.  The problem of extracting
$C_+$ and $C_-$ now becomes the problem of determining the inverse transition
temperature $\betat$.

\begin {figure}
\vbox
    {%
    \vspace* {-17pt}
    \begin {center}
        \leavevmode
        
        \epsfbox {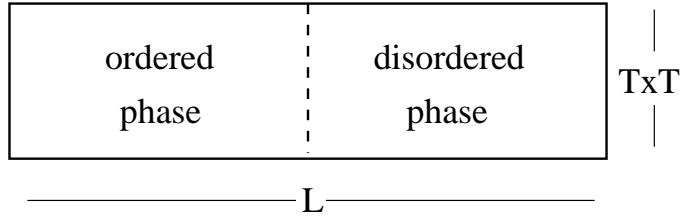}
    \end {center}
    \caption
        {%
        \label {fig:mixed-lattice}
        Initial conditions used on asymmetric lattices for determining
        the transition temperature.
        }%
    }%
\end {figure}

\begin {figure}
\vbox
    {%
    \vspace* {-17pt}
    \begin {center}
        \leavevmode
        
        \epsfbox [150 50 500 500] {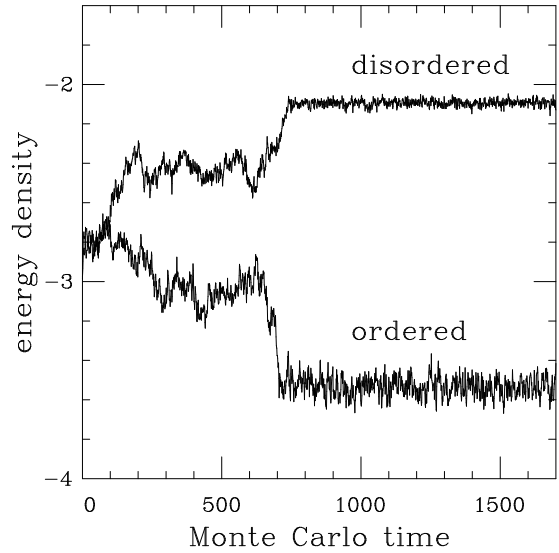}
    \end {center}
    \caption
        {%
        \label {fig:E_vs_tau}
        Examples of evolution of energy density with Monte Carlo time
        starting from the mixed phase of fig.~\protect\ref{fig:mixed-lattice}.
        The data is shown for $x{=}0.6$ on a $160\times40^2$ lattice near
        the transition temperature ($\beta=0.180263$).
        }%
    }%
\end {figure}

\begin {figure}
\vbox
    {%
    \vspace* {-17pt}
    \begin {center}
        \leavevmode
        
        \epsfbox [150 327 500 480] {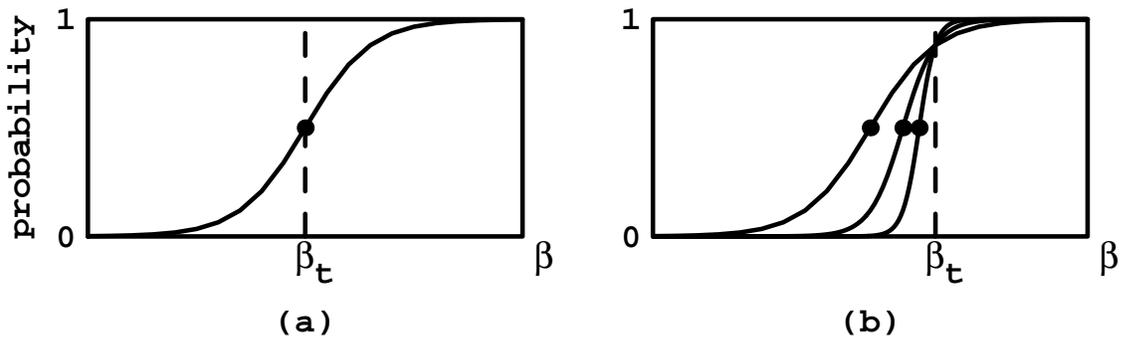}
    \end {center}
    \caption
        {%
        \label {fig:ruin}
        (a) Probability of ending in the ordered
        phase vs.\ $\beta$ when starting
        with mixed initial conditions; (b) the same but including finite $L$
        effects.  The narrower curves correspond to larger $L$.
        }%
    }%
\end {figure}

   We determine the transition temperature by working on asymmetric lattices
of size $L\times T\times T$ with $L\ge T$.
We divide the $L$ dimension in half,
and we start the lattice in the ordered phase in one half and the disordered
phase in the other, as depicted in fig.~\ref{fig:mixed-lattice}.
Next we evolve the system with
our Monte Carlo update procedure.  The domain walls will feel a net pressure
to expand the phase with the lowest free energy.
By tracking whether the system eventually evolves into the ordered or
disordered phase, and finding the value of $\beta$ where the favored phase
changes, one determines the transition temperature.  The story is a little more
complicated, however.  Exactly at the transition temperature, neither phase is
favored and the domain walls will random walk until, randomly, they collapse
the system into one phase or the other.
Fig.~\ref{fig:E_vs_tau} shows
an example from our simulations of collapsing into either
phase at the transition temperature.
Slightly away from the transition
temperature, there will be a slight bias to the random walks, but there will
still be some chance of ending in the disfavored phase.  The probability of
ending up in one particular phase, say the ordered one, therefore has a
dependence on temperature like that sketched in fig.~\ref{fig:ruin}a.
It can be modeled
as the solution to a classic problem from probability theory---the
Gambler's Ruin problem---which is described in
Appendix~\ref{app:gambler}.
Our technique for
determining $\betat$ is therefore to make multiple, independent runs at each
temperature, in order to numerically extract the curve of fig.~\ref{fig:ruin}a,
and then to
fit for $\betat$.  There is one more complication:
even if the transverse dimension $T$ is large
compared to the correlation length, there are systematic errors in this
procedure that have only power-law fall-off with increasing
longitudinal dimension
$L$, as depicted in fig.~\ref{fig:ruin}b.
To get the best estimate of the transition
temperature, we make the additional step of numerically extracting the
$L{\to}\infty$ limit.  A model for the finite $L$ corrections is discussed in
Appendix~\ref{app:gambler} and sec.~\ref{sec:beta}.

All of our calculations were carried out on a handful of SGI Indy R4600
workstations, taking roughly 2 to 45 mins.\ per energy decorrelation time
(depending on $x$) on an $80^3$ lattice at the transition.


\section {Analysis of Data and \lowercase{$x{\to}0$}}
\label{sec:analysis}

\begin {figure}
\setlength\unitlength{1 in}
\vbox
   {%
   \begin {center}
   \begin {picture}(5,7)(0.1,0.3)
      \put(-0.2,3.3){
        \leavevmode
        
        \epsfbox [120 25 500 450] {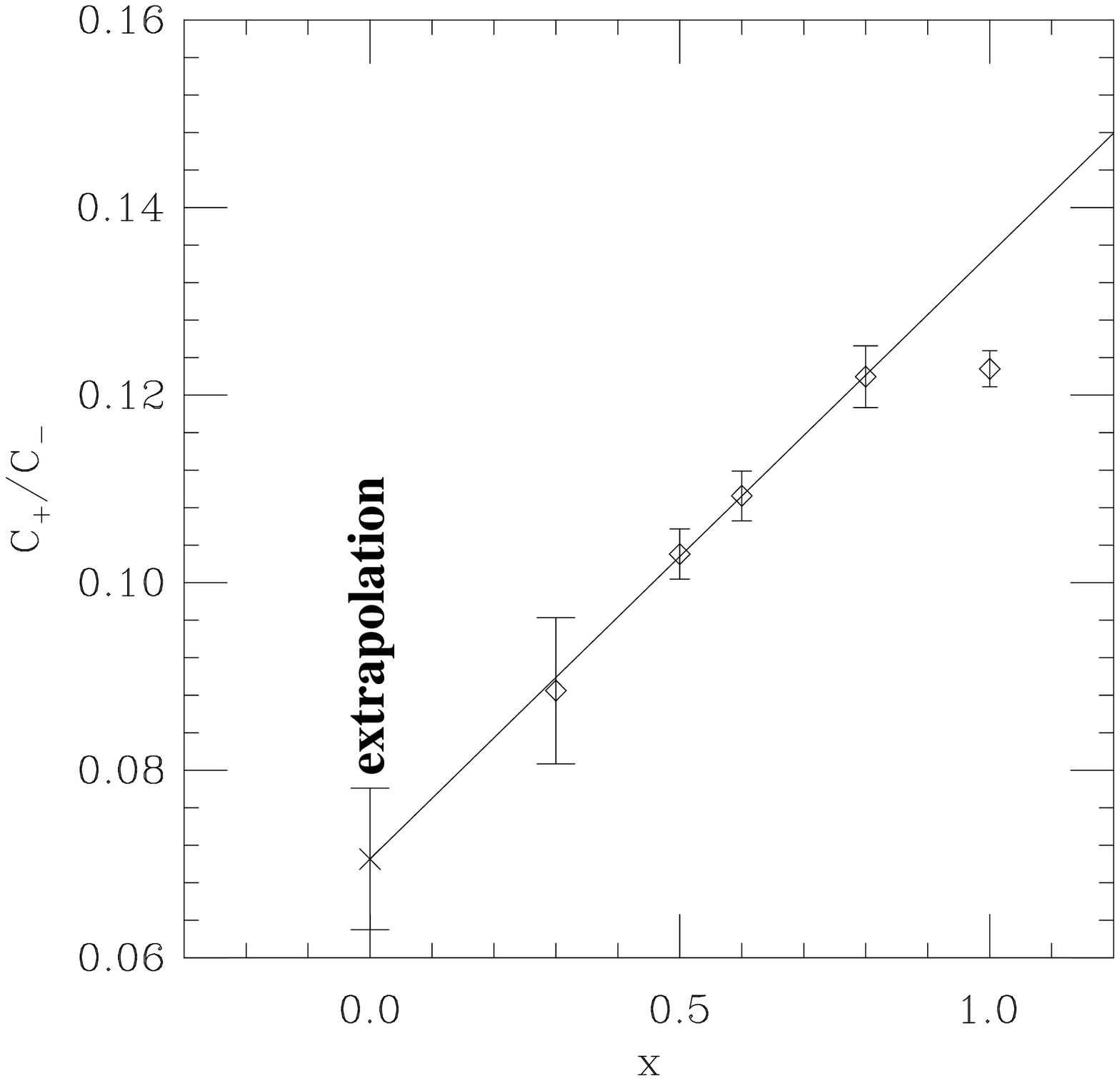}
      } 
      \put(3.2,3.3){
        \leavevmode
        
        \epsfbox [150 60 500 500] {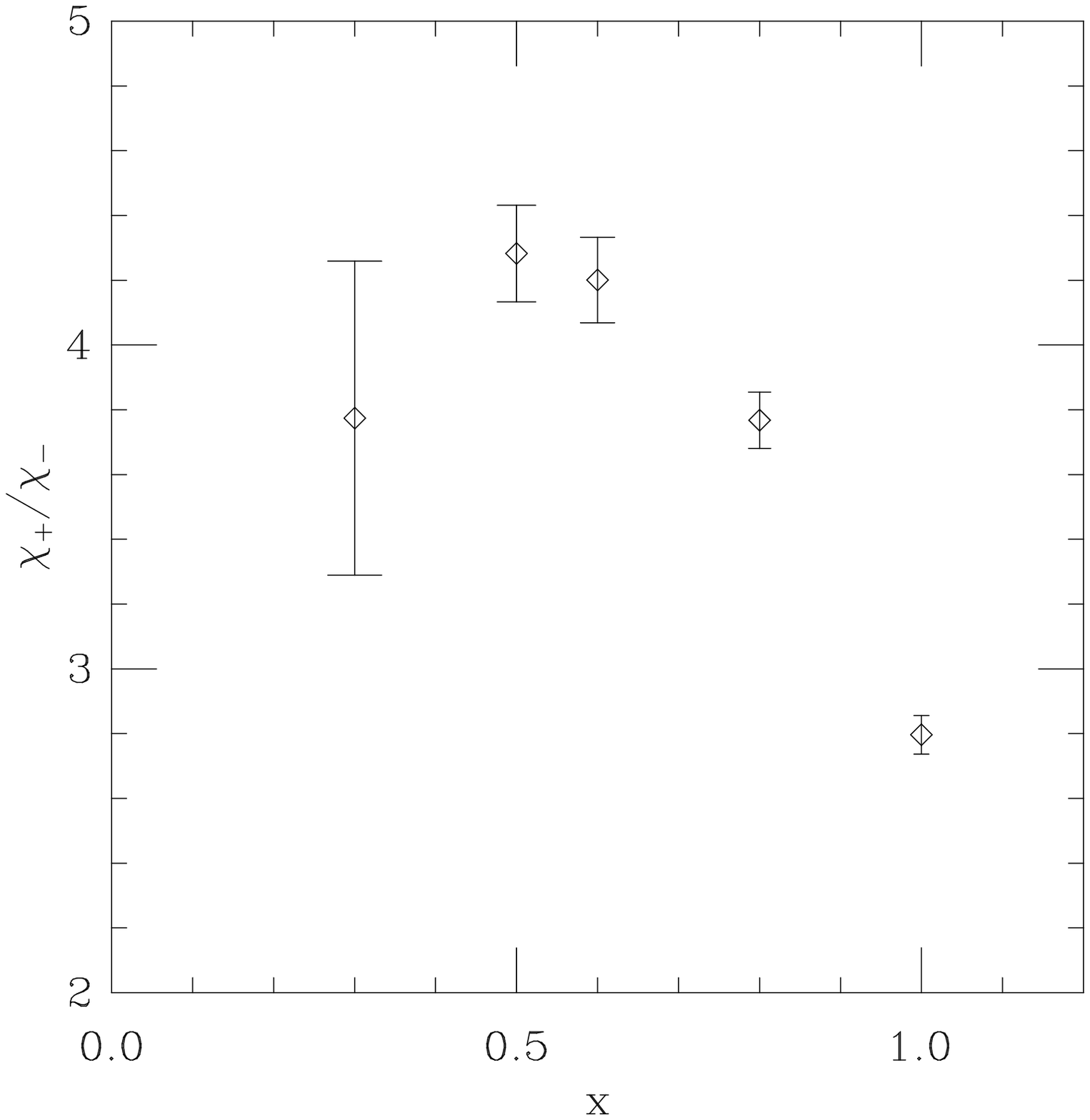}
      }
      \put(-0.2,5.8){(a)}
      \put(3.2,5.8){(b)}
      \put(-0.2,2.8){(c)}
      \put(3.2,2.8){(d)}
      \put(-0.2,0.3){
        \leavevmode
        
        \epsfbox [150 60 500 500] {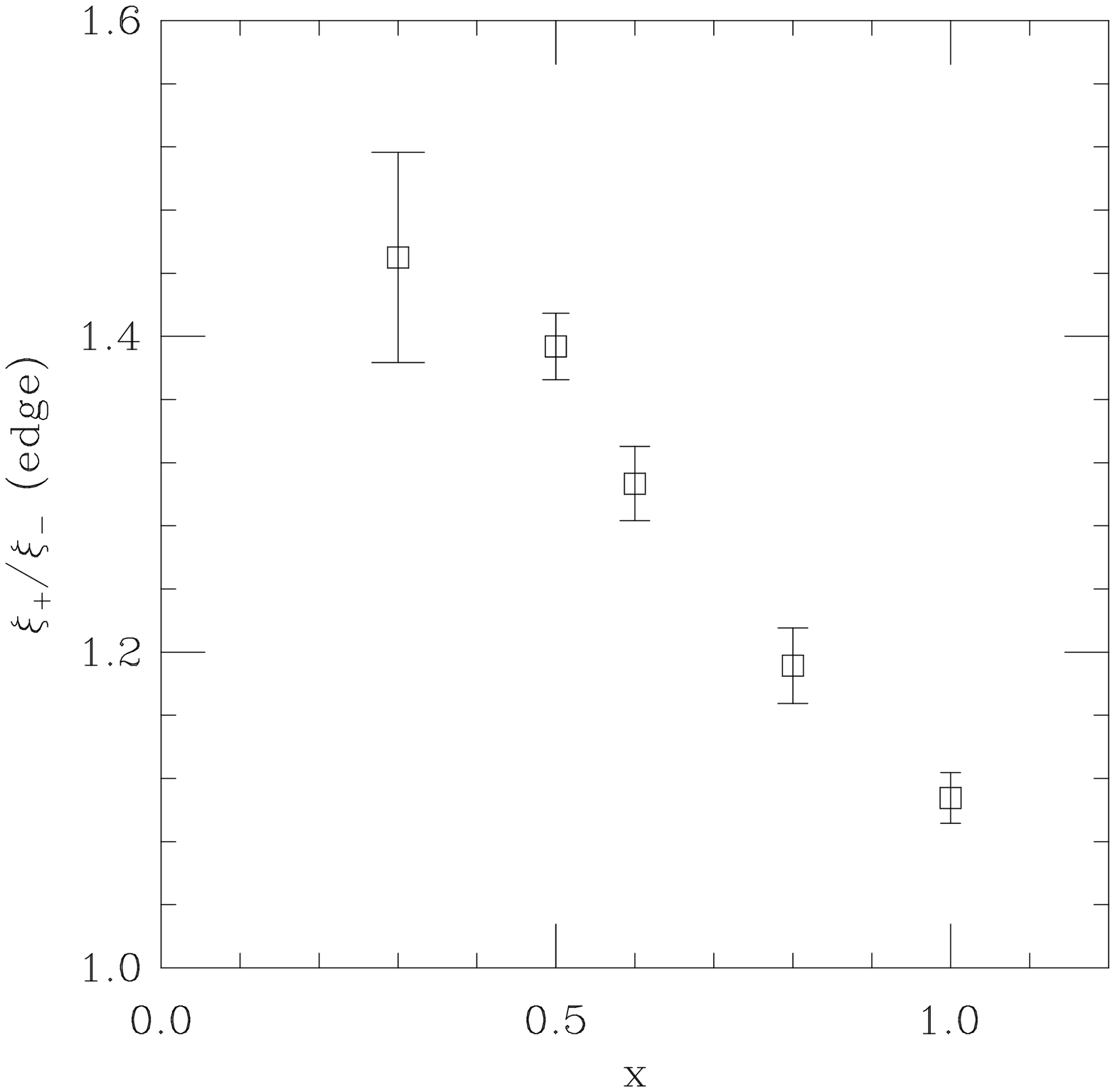}
      }
      \put(3.2,0.3){
        \leavevmode
        
        \epsfbox [150 60 500 500] {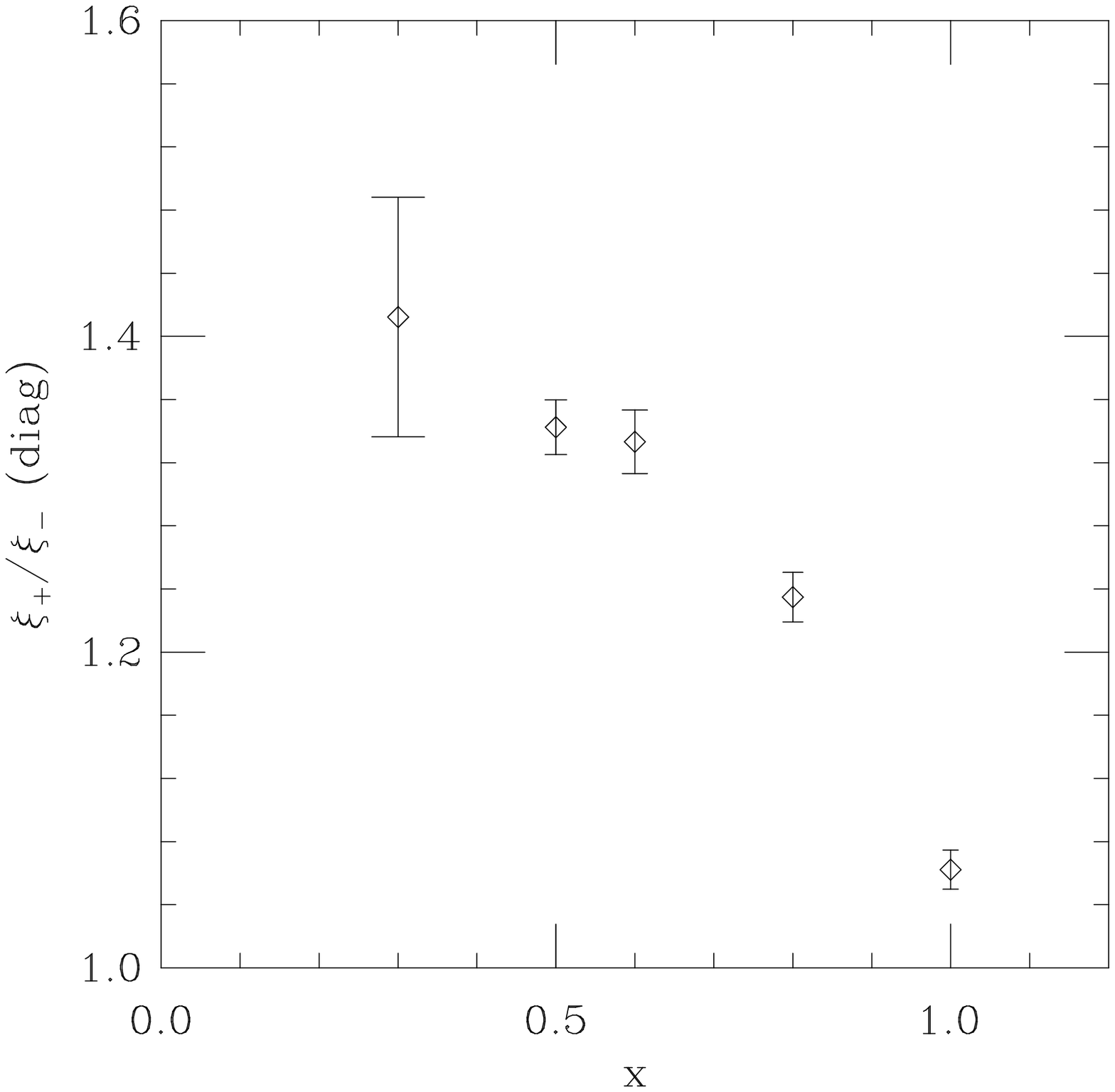}
      }
   \end {picture}
   \end {center}
   \caption
       {%
       Plots of the data in table~\protect\ref{tab:ratios}.
       For $C_+/C_-$, our best-fit interpolation and the extrapolated
       $x{\to}0$ value are also shown
       (see sec.~\protect\ref{sec:limit extraction}).
       Extrapolations of the other ratios are shown in
       fig.~\protect\ref{fig:ratios_vs_scale}.
       \label{fig:ratios_vs_x}
       }%
   }%
\end {figure}

Table~\ref{tab:ratios}
and fig.~\ref{fig:ratios_vs_x}
summarize our results for various ratios as a function
of $x$ on simple cubic lattices;
table~\ref{tab:data} shows in more detail the individual quantities
that went into determining the ratios.
We will explain in more detail our determination of these quantities
in sec.~\ref{sec:details}.
In this section, we will focus on understanding the
small $x$ behavior of the system and the extraction of the
$x\to0$ limits of the ratios.
The errors quoted for our data
are mostly statistical but include the systematic error
from our uncertainty in the transition temperatures $\betat$.
We estimate finite size effects to be no larger than our quoted
total errors except possibly for
our lowest value $x{=}0.3$ of $x$, where we do not have a very good
estimate.
For finite correlation length, the correlation length measured along
edges of the lattice need not be exactly the same as that measured
along diagonals, and so we have listed these cases separately.
The $x{\to}0^+$ ratio $\xi_+/\xi_-$ in the two cases, however, should
be the same.

\begin{table}
\def\cen#1{\multicolumn{1}{c}{$#1$}}
\def\cenb#1{\multicolumn{1}{c|}{$#1$}}
\begin {center}
\tabcolsep=6pt

\begin {tabular}{|lllll|}                                      \hline
$x$   &\cen{C_+/C_-}&\multicolumn{2}{c}{$\xi_+/\xi_-$}
                                      &\cenb{\chi_+/\chi_-} \\
      &             &\multicolumn{1}{c}{edge}
                    &\multicolumn{1}{c}{diag.} &                 \\ \hline
1.0   & 0.1228(19)  & 1.108(16)$^{\rm b}$
                    & 1.062(12)$^{\rm b}$   &  2.80(6)           \\
0.8   & 0.122(3)    & 1.191(24)$^{\rm b}$
                    & 1.235(16)$^{\rm b}$   &  3.77(9)           \\
0.6   & 0.109(3)    & 1.307(24)$^{\rm b}$
                    & 1.333(20)$^{\rm b}$   &  4.20(13)          \\
0.5   & 0.103(3)    & 1.394(21)$^{\rm b}$
                    & 1.342(17)$^{\rm b}$   &  4.28(15)          \\
0.3   & 0.088(8)$^{\rm a}$
                    & 1.45(7)$^{\rm a,b}$
                    & 1.41(8)$^{\rm a,b}$   & 3.8(5)$^{\rm a}$   \\ \hline
\end {tabular}
\end {center}
\caption
    {%
    \label {tab:ratios}
    Summary of relative discontinuities of specific heat, correlation length
    (along lattice edges and diagonals),
    and susceptibility vs. $x$.  Errors include statistical errors
    and systematic errors due to the uncertainty in the determination of
    transition temperatures.  Finite volume errors are not included and are
    estimated to be no larger than the total errors quoted above,
    except possibly for (a)
    the case $x{=}0.3$ where we do not have a good
    estimate of them.  (See sec.~\protect\ref{sec:details} for details.)
    (b) Errors for some correlation lengths are quite likely underestimated
    (see sec.~\protect\ref{sec:xi} for details).
    }%
\end{table}

\begin {figure}
\vbox
    {%
    \begin {center}
        \leavevmode
        
        \epsfbox [150 60 500 500] {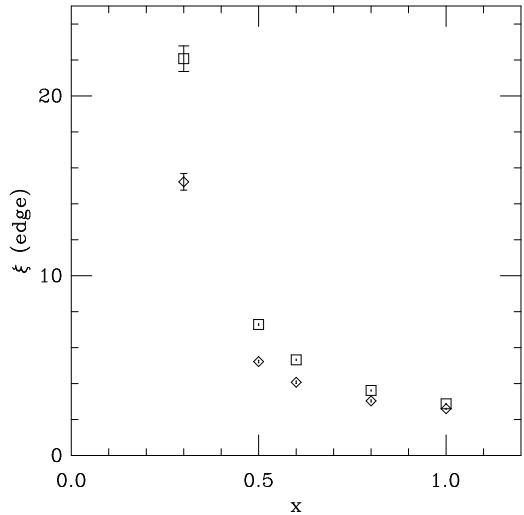}
    \end {center}
    \caption
        {%
        \label {fig:xi_vs_x}
          The correlation length $\xi^\edge$ (measured along lattice edges) at
          the transition vs.\ $x$ for the
          disordered phase (squares) and ordered phase (diamonds).
        }%
    }%
\end {figure}

Our lowest value $x=0.3$ of $x$ is small in the sense that
the transition is relatively weak: the correlation length at the transition
is order 20 (see table~\ref{tab:data}).  Indeed, it is precisely
the rapid growth of correlation length with decreasing $x$ shown in
fig.~\ref{fig:xi_vs_x}
(combined with the requirement that lattices be large compared
to $\xi$ to avoid transitions between the phases) that has prevented us
from simulating even smaller $x$.
However, despite the fact that $x=0.3$ is ``small,'' the quality
of our data in fig.~\ref{fig:ratios_vs_x}
is such that, when attempting to extract
the $x{\to}0$ limits, we would clearly benefit
tremendously from knowing {\it a priori} how the
correction to the $x\to0$ limit should scale for small $x$.

We shall first discuss how the individual quantities
$C_\pm$, $\xi_\pm$, and $\chi_\pm$ scale with $x$.
Then we turn to the scaling of the corrections to $x{\to}0$ limit
of ratios.  Finally, armed with this analysis, we shall fit the
data of fig.~\ref{fig:ratios_vs_x} as best we can.

\begin{table}
\def\cen#1{\multicolumn{1}{c}{$#1$}}
\def\cenb#1{\multicolumn{1}{c|}{$#1$}}
\begin {center}
\tabcolsep=6pt

\begin {tabular}{|llrrrrr|}                                              \hline
$x$   &\cen{\betat}   & \cen{\epsilon_+}
                                  &\cen{C_+}&\cen{\xi_+^\edge}
                                            &\cen{\xi_+^\diag}&\cenb{\chi_+}\\
      &               & \cen{\epsilon_-}
                                  &\cen{C_-}&\cen{\xi_-^\edge}
                                            &\cen{\xi_-^\diag}&\cenb{\chi_-}\\
                                                                         \hline
1     &  0.157154(4)  &-2.2734(3) &1.663(13)& 2.878(24)& 2.985(14)& 33.1(3)  \\
      &               &-4.6009(23)&13.54(18)& 2.60(3)  & 2.81(3)  & 11.84(22)\\
0.8   &  0.168149(3)  &-2.1552(4) &1.836(24)& 3.62(3)  & 3.74(3)  & 51.9(9)  \\
      &               &-4.1726(23)&15.1(4)  & 3.04(6)  & 3.03(3)  & 13.78(20)\\
0.6   &  0.180272(3)  &-2.0942(4) & 2.33(3) & 5.32(4)  & 5.47(5)  &111(3)    \\
      &               &-3.5359(23)& 21.3(4) & 4.07(7)  & 4.10(5)  & 26.3(3)  \\
0.5   &  0.186750(4)  &-2.0797(3) & 2.72(3) & 7.28(5)  & 7.29(4)  &203(3)    \\
      &               &-3.145(3)  &26.4(6)  & 5.23(7)  & 5.43(6)  & 47.3(1.3)\\
0.3   &  0.2003659(15)&-2.0658(4) &5.16(21) &22.1(7)   &21.6(8)   &1730(180) \\
      &               &-2.362(3)  &  58(4)  &15.2(5)   &15.3(6)   & 460(30)  \\
0     &  0.221652(3)$^{\rm a}$&   &         &          &          &          \\
\hline
\end {tabular}
\end {center}
\caption
    {%
    \label {tab:data}
    Summary of inverse transition temperature $\betat$, energy density
    $\epsilon_\pm$, specific heat density $C_\pm$, 
    correlation length $\xi_\pm$, and susceptibility $\chi_\pm$
    as a function of $x$ on a simple
    cubic lattice.  $+$ and $-$
    denote the disordered and ordered phases, respectively, at
    the transition temperature.
    For some $x$, the errors on the correlation lengths are likely
    underestimated (see~\protect\ref{sec:xi} for details).
    The $x{=}0$ Ising model $\betat$ (a) is taken from
    ref.~\protect\cite{baillie}.
    }%
\end{table}


\subsection {The Problem with Crossover Exponents}
\label{sec:crossover}

As we shall review below, one already knows how various dimensionful
quantities, such as the correlation lengths $\xi_\pm$, should diverge
as $x{\to}0$.  The divergence is characterized by crossover (or
``tricritical'') exponents, {\it e.g.}
\begin {equation}
   \xi_\pm \sim x^{-y}, \qquad x\to 0.
\end {equation}
In the case at hand, crossover exponents such as $y$ may be determined
from knowledge of Ising model critical exponents.
In the Ising model, the scaling dimension of the nearest-neighbor
interaction in the Hamiltonian density is
\begin {equation}
   \dim[ s_i s_{i+\hat e} ] = d - {1\over\nu} = {d\over2} - {\alpha\over2\nu}
   \,,
\end {equation}
where $d{=}3$ is the dimension of space.
Now consider the scaling dimension of the interaction
$(st)_i (st)_{i+\hat e}$ in the Ashkin-Teller model.
For the purpose of understanding the crossover behavior at very small $x$,
we can ignore the effect of the interaction itself on its scaling and
instead consider the scaling dimension of the operator in the
$x{=}0$ limit.  Then the operator factorizes:
\begin {equation}
   D \equiv
   \dim[ (st)_i (st)_{i+\hat e} ] = \dim[ s_i s_{i+\hat e} ]
                                    + \dim[ t_i t_{i+\hat e} ]
         = d - {\alpha\over\nu} \,.
\end{equation}
Deviation from Ising behavior will occur once this operator becomes
important, which in our case means that the scale $\xi$ of the first-order
transition is related to the interaction's coefficient $x$ by
\begin {equation}
   x \sim \xi^{D-d} \sim \xi^{-\alpha/\nu} \,.
\label{eq:x-xi}
\end {equation}
Using Ising scaling laws to get the power-law relationship of $\xi$ with
other quantities, one gets
\begin {mathletters}%
\label {eq:x-exponents}%
\begin {eqnarray}
   \xi_\pm  &\sim& x^{-\nu/\alpha} \sim x^{-5.7(3)}    \,,
\\
   \chi_\pm &\sim& x^{-\gamma/\alpha} \sim x^{-11.3(5)} \,,
\\
   C_\pm    &\sim& x^{-1} \,,
\end {eqnarray}%
\end{mathletters}%
where we have used the Ising model exponents \cite{nickel&rehr}
\begin {equation}
   \alpha \simeq 0.110(5)\,, \qquad
   \nu \simeq 0.6300(15)\,, \qquad
   \gamma \simeq 1.2405(15)\,.
\end {equation}

This result means, in the small $x$ limit, that the correlation length
and susceptibility should grow by factors of roughly 40--60
and 1500--3500 respectively when $x$ is reduced by a factor of two!
There is clearly
no sign of such strong $x$ dependence in the data of table~\ref{tab:data}
and fig.~\ref{fig:xi_vs_x}.
A natural question now arises: Does this discrepancy indicate
failure to reach the small $x$ region of the Ashkin-Teller model
and so make our $x\to0$ limits for ratios suspect?

To address this question, one must first note that the
origin of the strong $x$ dependence in (\ref{eq:x-exponents}a-b) is
the small value of the specific-heat exponent $\alpha$ in the Ising
model.  It is useful to formally consider what would have happened if
$\alpha$ were arbitrarily small, or even zero.
In Appendix~\ref{app:crossover}, we
consider a simple model for the renormalization group flow
where one can treat this explicitly.
In the case
$\alpha{=}0$, the relationship (\ref{eq:x-xi}) generically becomes
$x \sim 1/\ln(\xi)$, and one gets
\begin {mathletters}%
\label {eq:x-exponential}%
\begin {eqnarray}
   \xi_\pm  &\sim& e^{k\nu/x} \,,
\\
   \chi_\pm &\sim& e^{k\gamma/x} \,,
\\
   C_\pm    &\sim& x^{-1} \,,
\\
   {C_+\over C_-} &\sim& {\xi_+\over\xi_-} \sim {\chi_+\over\chi_-} \sim 1 \,,
\end {eqnarray}%
\end {mathletters}%
where $k$ is some constant.
For arbitrarily small but non-zero $\alpha$,
one finds exponential scaling (\ref{eq:x-exponential}a-b)
of $\xi$ and $\chi$ for moderately
small $x$, but the power-law scaling form (\ref{eq:x-exponents}a-b) takes
over when $x$ drops below
\begin {equation}
   x_1 \sim {\alpha k\over \ln(1/\alpha k)} \,.
\end {equation}
This threshold has no effect, however, on the extraction of the limits
of ratios such as $\xi_+/\xi_-$.
(See Appendix~\ref{app:crossover} for details.)
So, though the formal limit $x \ll O(\alpha)$
is required to see correct scaling of $\xi$ and $\chi$, the weaker limit
$x \ll O(1)$ is adequate for extracting the dimensionless ratios of
interest.  Our failure to see the correct scaling of $\xi$ and $\chi$
in our numerical data should not, therefore, cause concern.
Roughly, we expect that $x$ should be small enough for computing
ratios if
the correlation length is large compared to the lattice spacing.

\begin {figure}
\vbox
    {%
    \begin {center}
        \leavevmode
        
        \epsfbox [150 60 500 500] {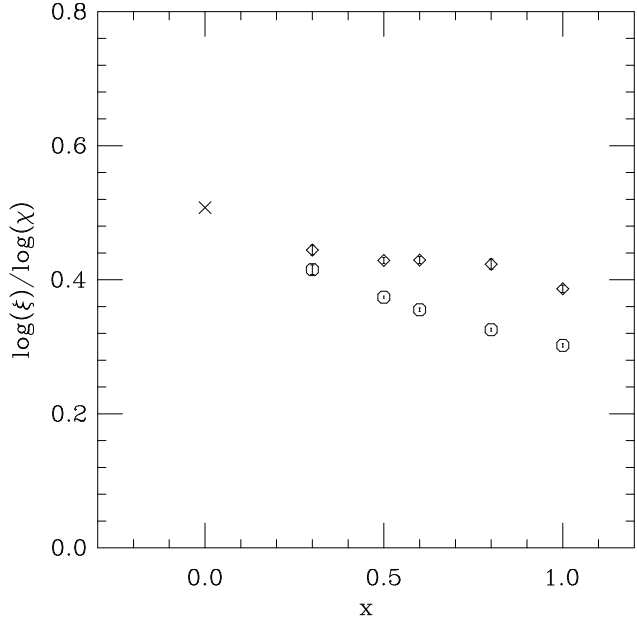}
    \end {center}
    \caption
        {%
        \label {fig:lnxi_lnchi}
          A test of scaling:
          $\ln(\xi^\edge)/\ln(\chi)$ vs.\ x for the ordered phase (diamonds)
          and disordered phase (circles).  The cross marks the theoretical
          value $\nu/\gamma$ for the $x{\to}0$ limit.
          The corresponding graph for $\ln(\xi^\diag)/\ln(\chi)$ is
          visually almost identical.
        }%
    }%
\end {figure}

\begin {figure}
\setlength\unitlength{1 in}
\vbox
   {%
   \begin {center}
   \begin {picture}(5,3.5)(0.1,0.3)
      \put(-0.2,0.3){
        \leavevmode
        
        \epsfbox [150 60 500 500] {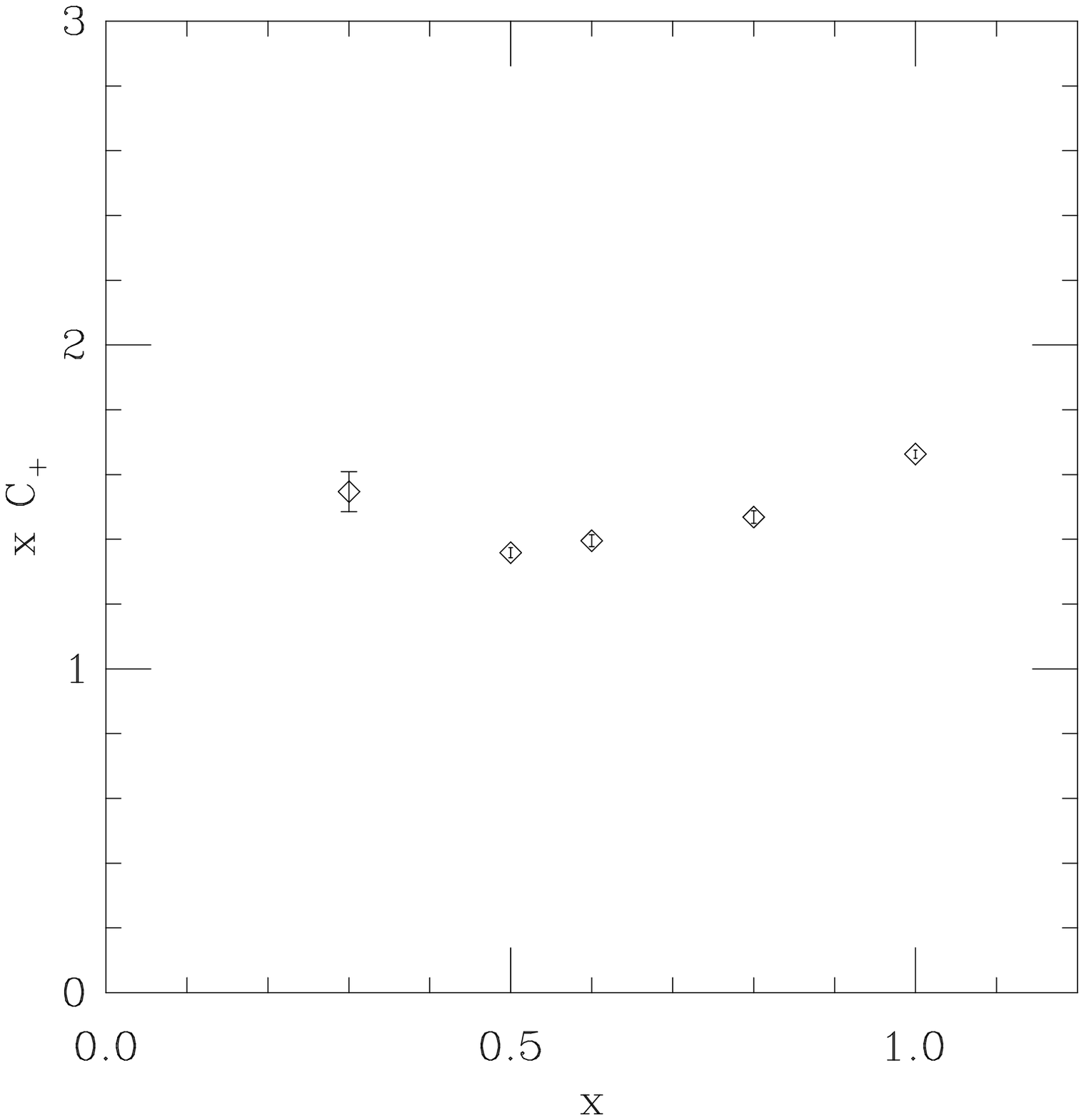}
      } 
      \put(3.2,0.3){
        \leavevmode
        
        \epsfbox [150 60 500 500] {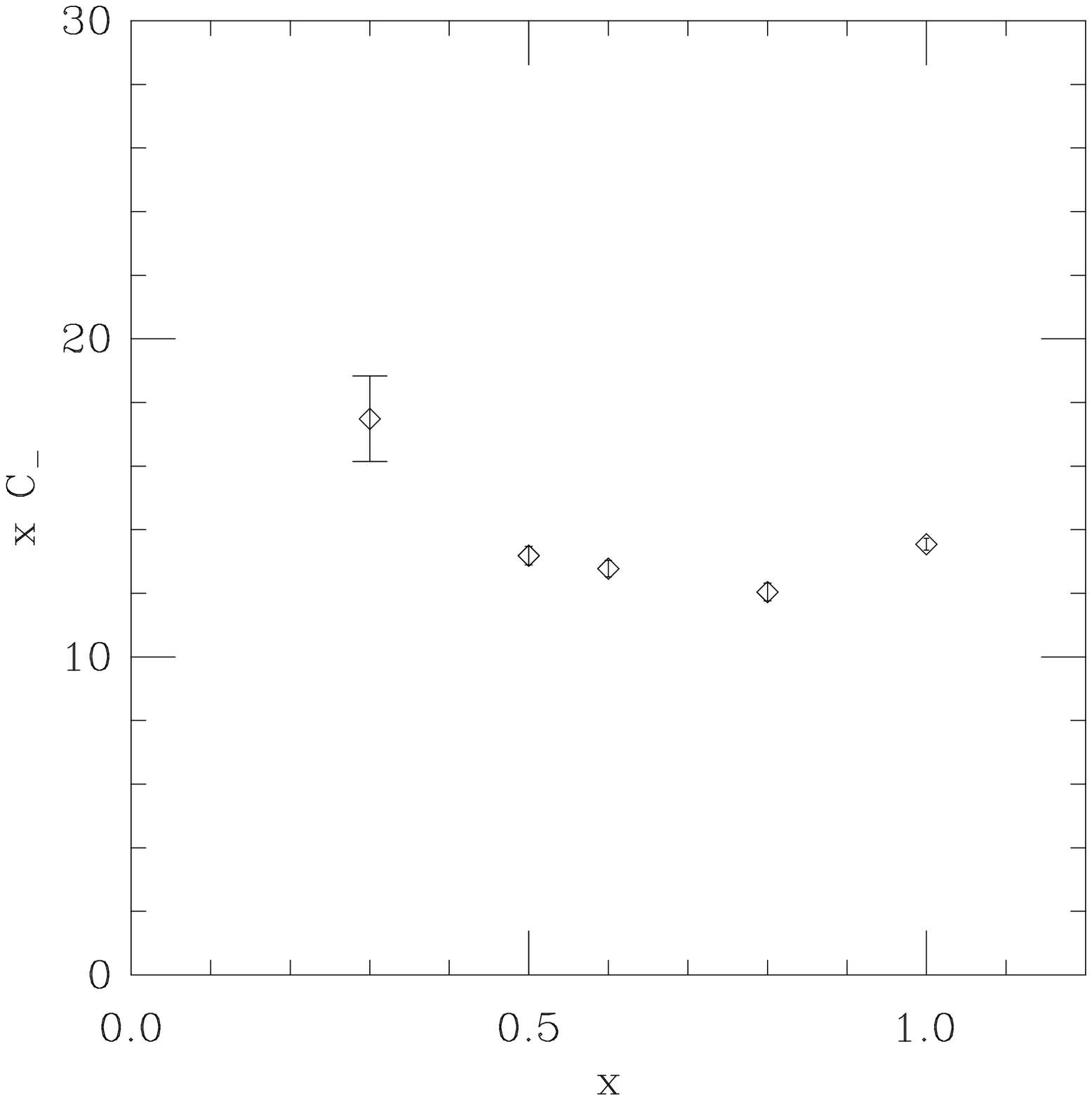}
      }
      \put(-0.2,2.8){(a)}
      \put(3.2,2.8){(b)}
   \end {picture}
   \end {center}
   \caption
       {%
       A test of scaling: (a) $x\,C_+$ and (b) $x\,C_-$ vs.\ x.
       \label{fig:cx_vs_x}
       }%
   }%
\end {figure}

We can test the above assertions by examining relationships that don't
depend on $\alpha$ and which should be satisfied for both the different
scaling behaviors (\ref{eq:x-exponents}) and (\ref{eq:x-exponential}).
One such relationship is that $\xi_\pm \sim \chi_\pm^{\nu/\gamma}$;
therefore
\begin {equation}
   \lim_{x\to 0^+} {\ln\xi_\pm \over \ln\chi_\pm} = {\nu\over\gamma} \,.
\label{eq:lnlnlimit}
\end {equation}
Fig.~\ref{fig:lnxi_lnchi}
shows this ratio of logarithms for our data and the theoretical
limit (\ref{eq:lnlnlimit}).  The data appears consistent with the limit.
Another test is to check whether
$x \, C_\pm$ approaches a constant as $x{\to}0$.
Fig.~\ref{fig:cx_vs_x}a
shows $x \, C_+$ for our data, which looks reasonably good.
Fig.~\ref{fig:cx_vs_x}b
shows $x \, C_-$, which is reasonable except that the lowest
$x$ point is a bit high.
(We shall see below that the approach to the $x{\to}0$ limit
should be linear at small $x$.)

\subsection{Corrections to Scaling}

   In order to extrapolate $x{\to}0$ limits for our ratios in
fig.~\ref{fig:ratios_vs_x}, it helps tremendously to know how the corrections
to those limits scale with $x$.
For simplicity, let us first ignore the small $\alpha$ issue.
We have discussed in the previous
section how, for very small $x$, the system has Ising behavior
for distance scales up to order $\xi \sim x^{-\nu/\alpha}$.
Corrections to scaling arise because there are irrelevant
operators which have not quite scaled away for finite (but large)
$\xi$.
However, since the scaling of operators as $\xi \to \infty$
is determined by Ising behavior, we can extract the dimension of
the most important such operator from well-known results in the
Ising model: relative corrections to scaling behavior scale as
$\xi^{-\omega}$, where $\omega$ is the Ising model exponent
\cite{nickel&rehr}
\begin {equation}
   \omega \simeq 0.79(3) \,.
\label{eq:omega}
\end {equation}
Translating $\xi$ to $x$ using (\ref{eq:x-xi}), we then have, for example,
\begin {eqnarray}
   \chi_{\pm} &\sim& x^{-\gamma/\alpha} \, [1 + O(x^{\omega\nu/\alpha})] \,,
\\
   {\chi_+\over\chi_-} &\sim& O(1) + O(x^{\omega\nu/\alpha}) \,.
\end {eqnarray}
Similarly,
\begin {equation}
   {\xi_+\over\xi_-} \sim O(1) + O(x^{\omega\nu/\alpha}) \,.
\end {equation}
The specific heat ratio is slightly different.  In addition to the
divergent $O(\xi^{\alpha/\nu}) = O(x^{-1})$
contribution from long-distance modes, the specific
heat receives a direct contribution from short-distance modes that is
analytic in $\xi$ and is therefore $O(\xi^0) = O(x^0)$.  This is much more
important than the relative contribution discussed above and gives
\begin {eqnarray}
   C_{\pm} &\sim& O(\xi^{\alpha/\nu}) + O(\xi^0) \,,
\\
   {C_+\over C_-} &\sim& O(1) + O(x) \,.
\label{eq:C corrections}
\end {eqnarray} 

So far, we have discussed the correction to scaling in the limit of
arbitrarily small $x$ where everything scales with the correct Ising
model exponents.  But, as discussed in the previous section, the
values of $x$ we actually simulate are not small enough to reproduce
relationships involving the small Ising exponent $\alpha$.
The scaling (\ref{eq:C corrections}) of corrections to
$C_+/C_-$ is nonetheless in good shape because it does not depend
on $\alpha$ and would hold even if $\alpha$ were zero.
The scaling of corrections to other ratios can also be cast in a form
that does not depend on $\alpha$:
\begin {mathletters}%
\label{eq:other corrections}%
\begin {eqnarray}
   {\chi_+\over\chi_-} &\sim& O(1) + O(\chi_{\pm}^{-\omega\nu/\gamma}) \,.
\\
   {\xi_+\over\xi_-} &\sim& O(1) + O(\xi_{\pm}^{-\omega}) \,.
\end {eqnarray}
\end {mathletters}%
(\ref{eq:C corrections}) and (\ref{eq:other corrections}) are the forms
we fit our data to in order to extract the $x{\to}0$ limits of the ratios.


\subsection {Extraction of $x{\to}0$ limits from data}
\label{sec:limit extraction}

\begin {figure}
\setlength\unitlength{1 in}
\vbox
   {%
   \begin {center}
   \begin {picture}(5,7)(-0.2,0.3)
      \put(1.5,3.3){
        \leavevmode
        
        \epsfbox [120 25 500 450] {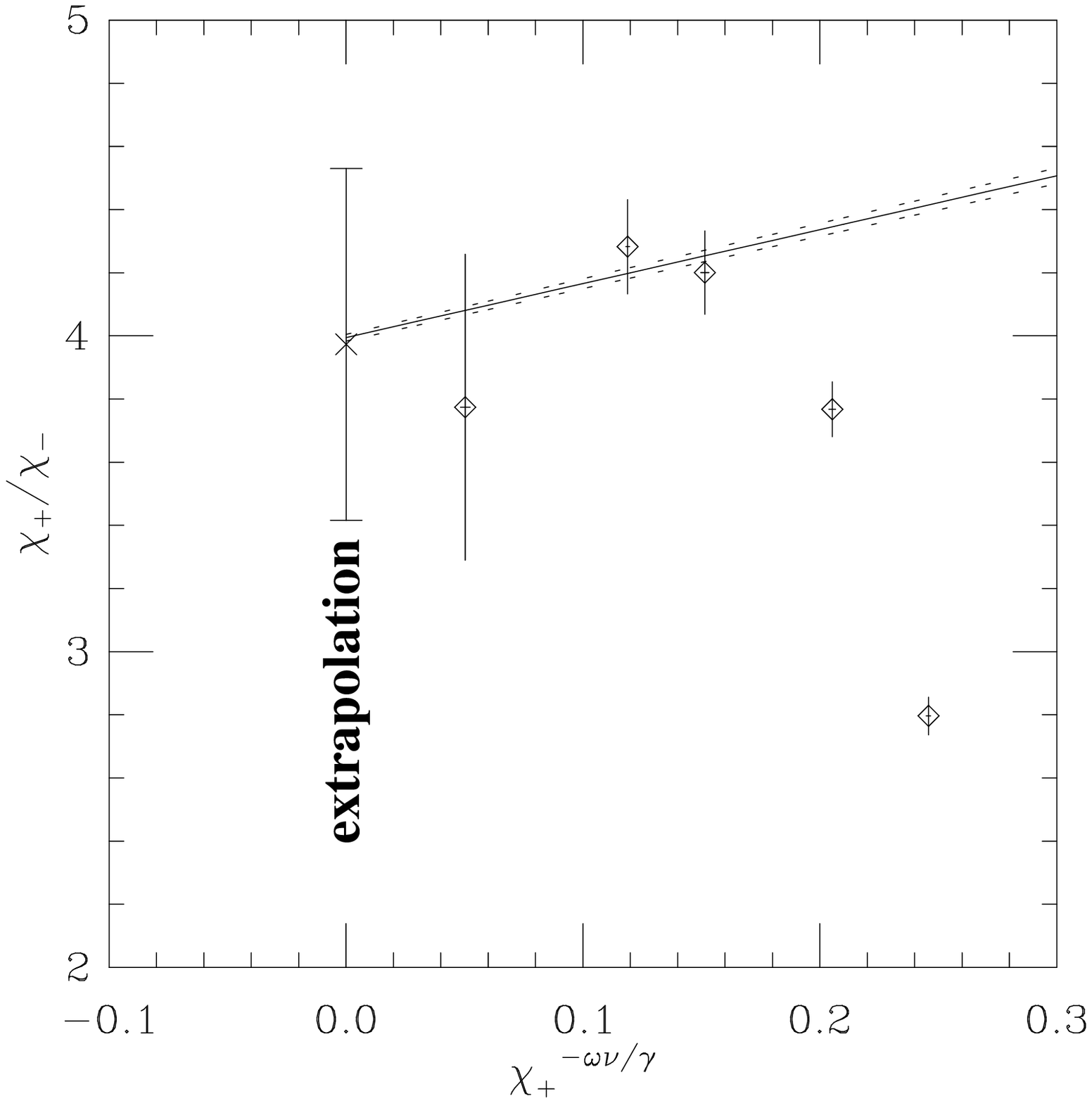}
      } 
      \put(1.5,5.8){(a)}
      \put(-0.2,2.8){(b)}
      \put(2.9,2.8){(c)}
      \put(-0.2,0.3){
        \leavevmode
        
        \epsfbox [120 25 500 450] {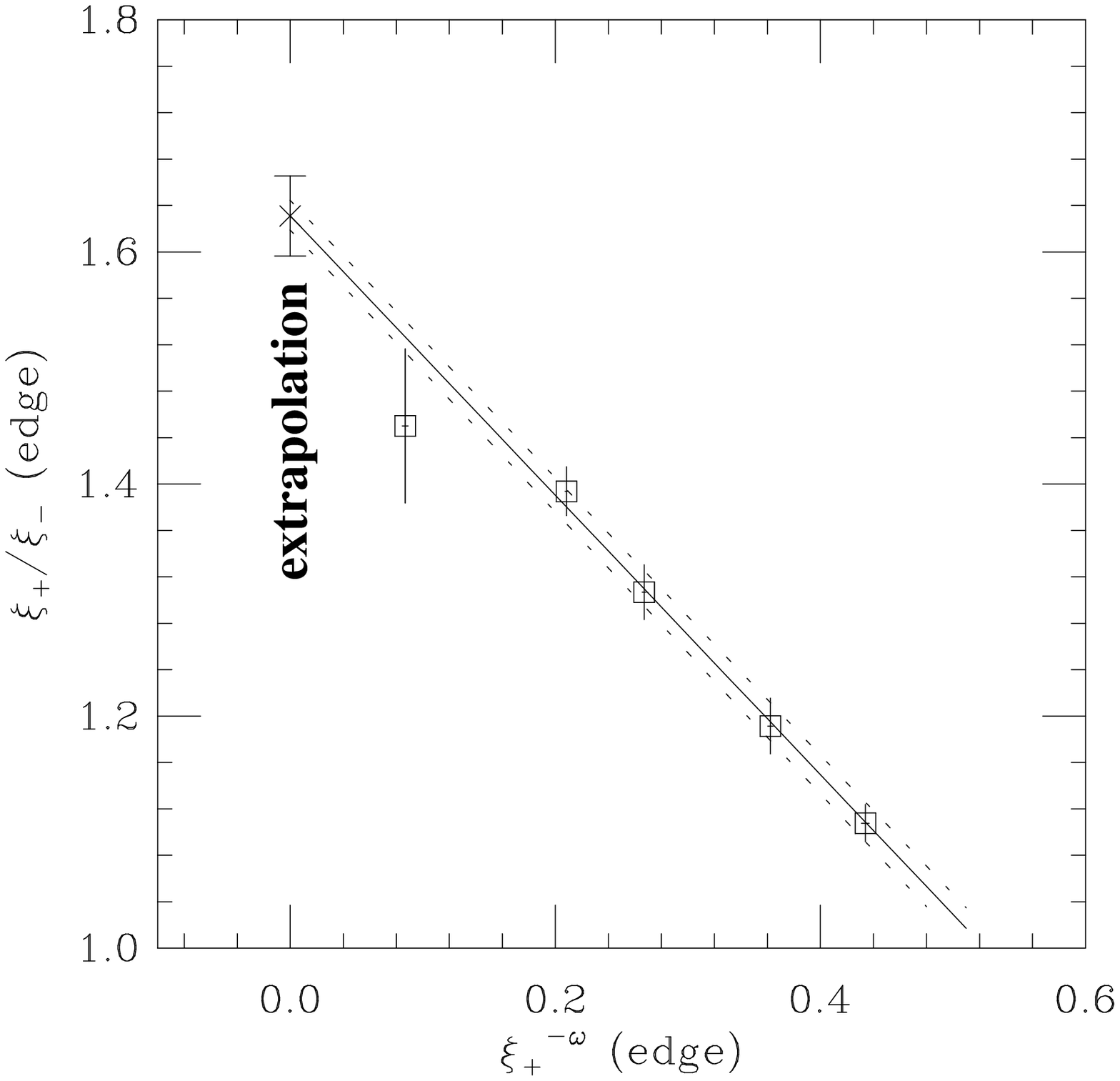}
      }
      \put(2.8,0.3){
        \leavevmode
        
        \epsfbox [105 25 485 450] {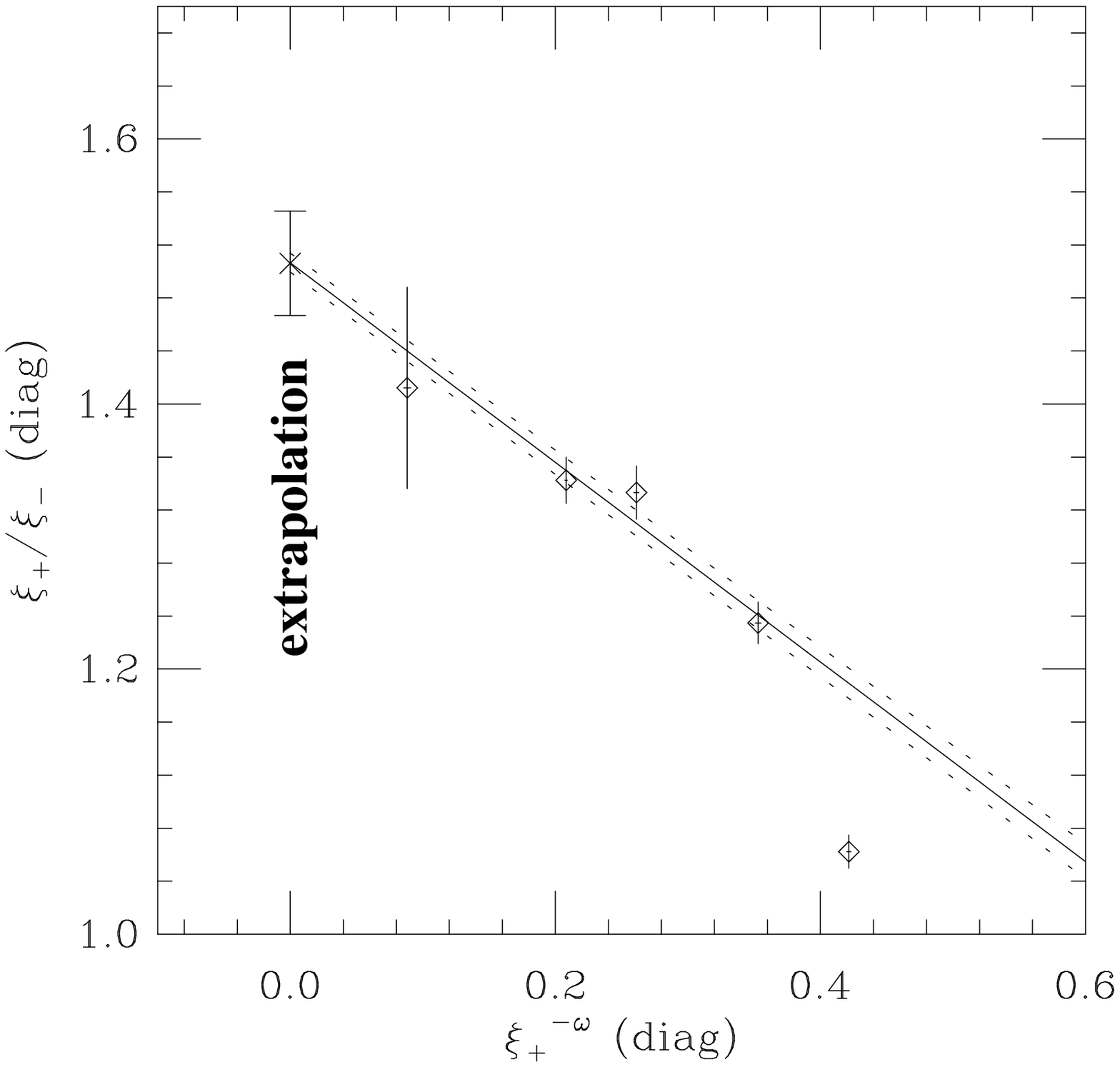}
      }
   \end {picture}
   \end {center}
   \caption
       {%
         Ratios vs.\ an appropriate scaling variable for corrections
         to the $x{\to}0$ limit:
         (a) $\chi_+/\chi_-$ vs.\ $\chi_+^{-\omega\nu/\gamma}$, and
         (b,c) $\xi_+/\xi_-$ vs.\ $\xi_+^{-\omega}$ for edges and diagonals.
         Statistical uncertainties in the ordinates of the data points
         are too small to be seen.  We have not shown for each data point the
         larger systematic uncertainties due to uncertainties in the
         Ising exponents (dominated by $\omega$).
         The solid line is our best fit, and the $x{=}0$
         value is our extrapolation.  The dotted lines show how our best
         fit changes as $\omega$ is varied over the uncertainty indicated
         in (\protect\ref{eq:omega}), but this change is negligible
         in (a).
         \label{fig:ratios_vs_scale}
       }%
   }%
\end {figure}

Now that we know how the corrections are supposed to scale, we can
attempt to fit our data to the appropriate form.  $C_+/C_-$ is already
plotted against the correct variable $x$ in fig.~\ref{fig:ratios_vs_x}
for a linear fit.  In fig.~\ref{fig:ratios_vs_scale}, we show
$\chi_+/\chi_-$ and $\xi_+/\xi_-$ plotted against $\chi_+^{-\omega\nu/\gamma}$
and $\xi_+^{-\omega}$ respectively, so that the fit should again be linear.
Our procedure is to fit the largest set of points, working from
the lowest $x$ up, that yields a reasonable chi-squared.

\begin {figure}
\setlength\unitlength{1 in}
\vbox
   {%
   \begin {center}
   \begin {picture}(5,4)(-0.2,0.3)
      \put(-0.2,2.8){(a)}
      \put(2.8,2.8){(b)}
      \put(-0.2,0.3){
        \leavevmode
        
        \epsfbox [120 25 500 450] {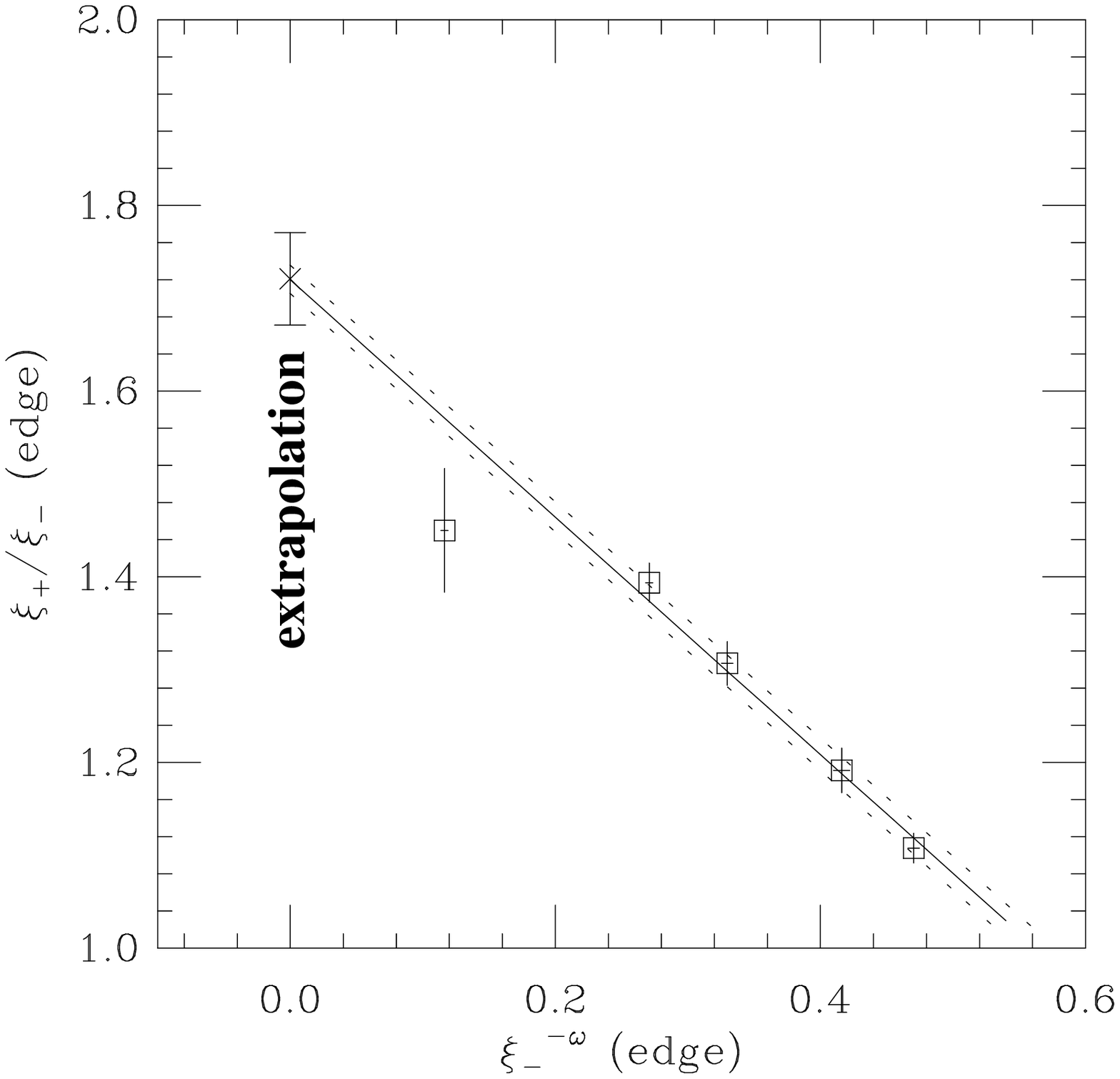}
      }
      \put(2.8,0.3){
        \leavevmode
        
        \epsfbox [120 25 500 450] {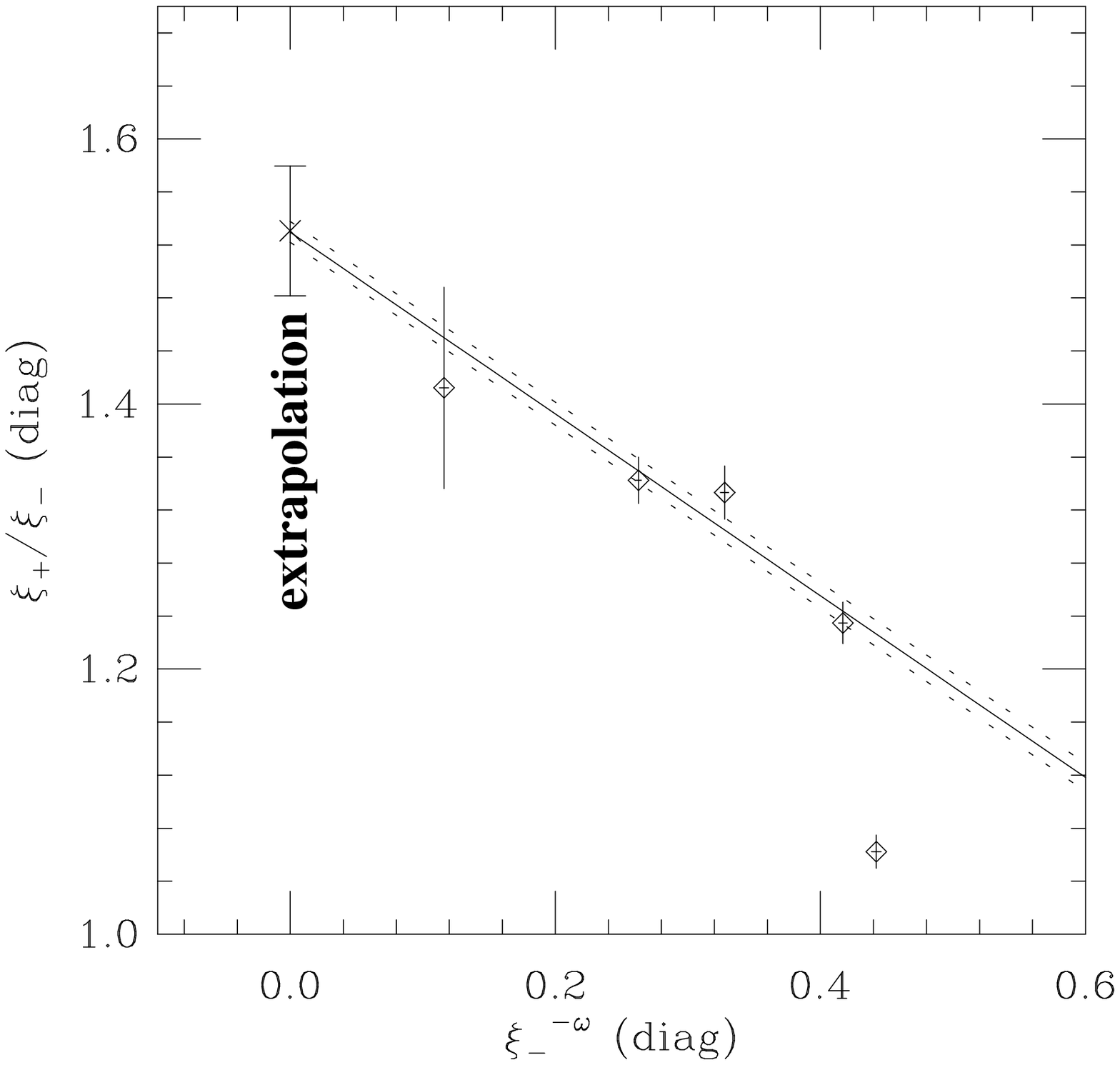}
      }
   \end {picture}
   \end {center}
   \caption
       {%
         $\xi$ ratios vs.\ $\xi_-^{-\omega}$ instead of $\xi_+^{-\omega}$.
         \label{fig:xiratios_vs_scale-}
       }%
   }%
\end {figure}

For $C_+/C_-$, fitting the points $x \le 0.8$ yields a freakishly high
confidence level of 98\% and
produces our
final result, eq.~(\ref{eq:cratio_result}), for the $x{\to}0$ limit.
Adding the point $x=1.0$ decreases the confidence level to
18\% and would change the limit to 0.084(4).
Because of its large uncertainty, our $x{=}0.3$ value
has an almost negligible effect on the fit and our final result.

Next consider $\chi_+/\chi_-$.
Fitting the three points $x \le 0.6$ yields the best fit line
(34\% confidence level)
shown in fig.~\ref{fig:ratios_vs_scale}a and our final result,
eq.~(\ref{eq:chiratio_result}), for the $x{\to}0$ limit.
The systematic uncertainty in the values of the Ising critical
exponents is included in our error estimate.
Attempting to add the $x=0.8$ point gives
$\lim(\chi_+/\chi_-) = 4.8(3)$ with a fairly small confidence level
of 11\%.
We have parameterized corrections (\ref{eq:other corrections}) to scaling
by $\chi_+^{-\omega\nu/\gamma}$.
An alternative choice would be $\chi_-^{-\omega\nu/\gamma}$,
and we have used $\chi_+^{-\omega\nu/\gamma}$ simply because it
is more straightforward to measure.
If the fit is made against $\chi_-^{-\omega\nu/\gamma}$ instead of
$\chi_+^{-\omega\nu/\gamma}$, the result for the limiting ratio is 4.0(5)
and is consistent with the previous result.
In any case, the situation is somewhat
unsatisfactory to the extent that (1) we have fit only three points,
and (2) one of those points is our lowest $x$ value, $x=0.3$, where we
are less confident about finite volume errors
(see sec.~\ref{sec:volume}).

In sec.~\ref{sec:xi}, we explain that our errors on extracting the
correlation lengths may be underestimated.  Our procedure for estimating our
final error will simply be to interpolate the $x{\to}0$ ratio
in different ways that should in principle be equivalent.
For $\xi_+^\edge/\xi_-^\edge$, we can obtain a
reasonable (56\% confidence level) fit vs.\ $(\xi_+^\edge)^{-\omega}$
with all our data points, shown in fig.~\ref{fig:ratios_vs_scale}b;
for the diagonal ratio,
we need to drop the $x{=}1.0$ point from the fit
(improving the confidence level from $O(10^{-4}\%)$ to 36\%).
The results are
\begin {mathletters}%
\label{eq:xifit}%
\begin {equation}
   \lim_{x\to0^+} \> {\xi_+^\edge\over \xi_-^\edge} = 1.64(3) \,,
   \qquad
   \lim_{x\to0^+} \> {\xi_+^\diag\over \xi_-^\diag} = 1.51(4) \,.
\label {eq:xifit 1}
\end {equation}%
Alternate interpolations using $\xi_-^{-\omega}$ are shown in
fig.~\ref{fig:xiratios_vs_scale-}
and give
\begin {equation}
   \lim_{x\to0^+} \> {\xi_+^\edge\over \xi_-^\edge} = 1.72(5) \,,
   \qquad
   \lim_{x\to0^+} \> {\xi_+^\diag\over \xi_-^\diag} = 1.53(5) \,.
\label{eq:xifit 2}
\end {equation}%
\end {mathletters}%
The wide discrepancy of values in (\ref{eq:xifit 1}) and (\ref{eq:xifit 2})
should be taken as a
reflection of our systematic errors in
determining the correlation lengths and in extracting the $x{\to}0$ limit.
Our final result (\ref{eq:xiratio_result}) has been chosen to span all of
the above extrapolations.

Clearly it would be useful to have more good-quality data at small $x$,
and, in particular, $x=0.4$ suggests itself as a good candidate for
future simulation.  We estimate that $x=0.4$ would take us 2-3 CPU
years on our SGI Indys, and we have not yet attacked it.  The most
time-consuming part of the task is an accurate determination of the
transition temperature.


\section{Details}
\label{sec:details}

\subsection {Susceptibilities}

The definition (\ref{eq:chi-def}) of the susceptibility is equivalent to
\begin {equation}
   \chi = {1\over2N}
             \left[ \langle \tilde {\bf S}(0) \cdot
                              \tilde{\bf S}(0) \rangle_{{\bf h}=0}
            - \langle \tilde {\bf S}(0) \rangle^2_{\bf h\to0} \right]
\label{eq:chi1}
\end {equation}
or
\begin {equation}
   \chi = {1\over2N} \lim_{{\bf p}\to0}
         \langle \tilde {\bf S}({\bf p})^* \cdot \tilde {\bf S}({\bf p})
                    \rangle_{{\bf h}=0}
        \equiv \lim_{{\bf p}\to 0} \chi({\bf p}) \,,
\label{eq:chi2}
\end {equation}
where $N$ is the number of lattice sites and $\tilde {\bf S}({\bf p})$
is the Fourier transform of the spin fields ${\bf S}_i = (s_i,t_i)$:
\begin {equation}
   \tilde {\bf S}({\bf p}) \equiv
       \sum_{\bf x} {\bf S}({\bf x}) e^{i {\bf p}\cdot{\bf x}}
   \,.
\end {equation}
To measure the disordered-phase susceptibility in our simulations,
we simply use the first term of (\ref{eq:chi1}).

\begin {figure}
\setlength\unitlength{1 in}
\vbox
   {%
   \begin {center}
   \begin {picture}(5,3.5)(0.1,0.3)
      \put(-0.2,0.3){
        \leavevmode
        
        \epsfbox [130 40 500 500] {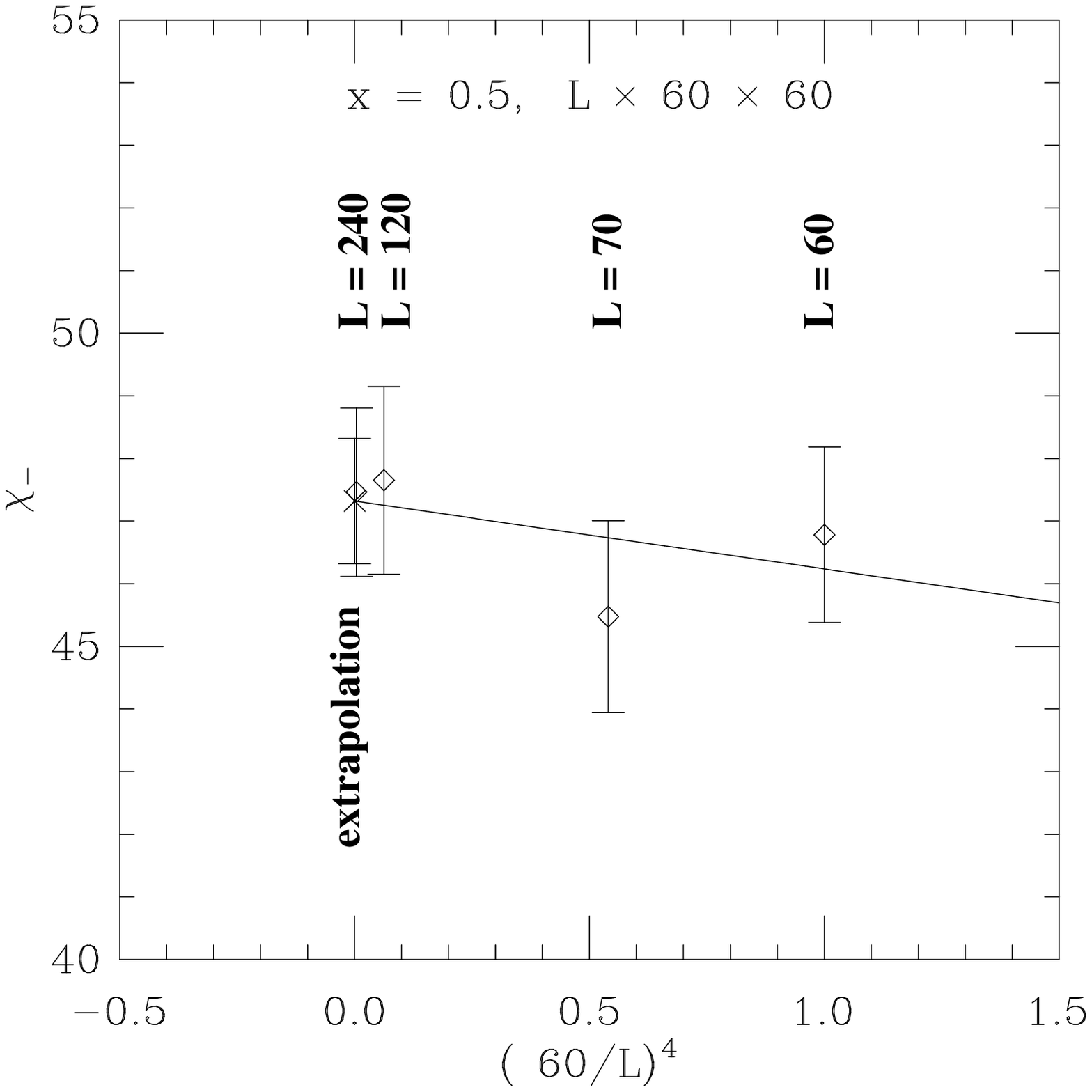}
      } 
      \put(3.2,0.3){
        \leavevmode
        
        \epsfbox [130 40 500 500] {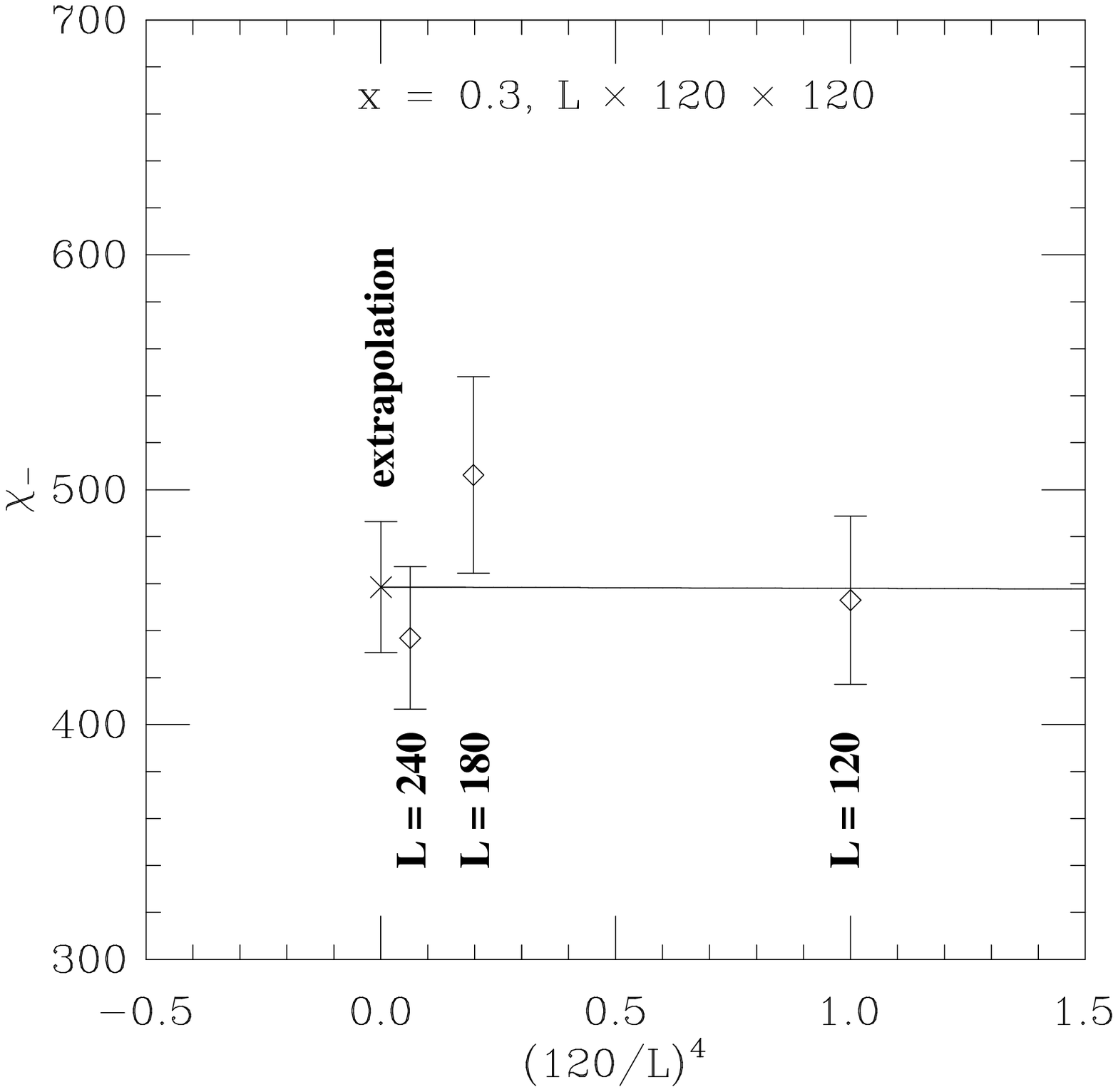}
      }
      \put(-0.2,2.5){(a)}
      \put(3.2,2.5){(b)}
   \end {picture}
   \end {center}
   \caption
       {%
         Measurements of $\chi_-$ on $L\times T\times T$ lattices, using the
         interpolation (\protect\ref{eq:chi2}), vs.\ L.  The solid line
         is the best fit of the results to $a + b L^{-4}$, and the
         cross (and smaller error bar)
         at $L{=}\infty$ shows the extrapolation from this fit.
         Two cases typical of our data are shown:
         (a) $x{=}0.5$, $T{=}60$, with 63\% confidence level;
         (b) $x{=}0.3$, $T{=}120$, with 18\% confidence level.
         \label{fig:chim_vs_L}
       }%
   }%
\end {figure}

For the ordered phase,
the subtraction in (\ref{eq:chi1}) is delicate and we instead use
(\ref{eq:chi2}), taking long, asymmetric lattices of size $L\times T\times T$
to get small values of the lowest non-zero momentum $p_{\rm min} = 2\pi/L$.
In more detail, we find we can get good estimates of
$\chi_-$ for fixed $L$ by measuring $\chi(p)$ for the two lowest
non-zero momenta in the long direction, $2\pi/L$ and $4\pi/L$, and
then extracting $\chi_-$ by fitting to the form
\begin {equation}
   \chi_-(p) \simeq {1\over\chi_-^{-1} + c \, p^2} \,.
\label{eq:chi12}
\end {equation}
These estimates for $\chi_-$ converge fairly quickly as $L$ is increased.
Typical examples of such dependence are shown in
fig.~\ref{fig:chim_vs_L} for $x{=}0.5$ and $0.3$.
For the sake of using all our data, our final results for $\chi_-$ in
table~\ref{tab:data} are the result of a fit of these results for
individual $L$ to the form%
\footnote{
  From (\ref{eq:chi12}), the difference between $\chi_-(p_{\rm min})$
  and the true susceptibility $\chi_-$ scales as $p_{\rm min}^2 \sim L^{-2}$
  for large $L$.  Interpolating $\chi(p)$ from the two lowest non-zero
  momenta improves the error to $L^{-4}$.  This is the reason for
  the form of our fit of these interpolated values vs.\ $L$.
}
$a + b L^{-4}$.
There is not too much
difference between this and the individual result for the largest $L$.
A list of the largest lattice size we use for each $x$ may be found
in table~\ref{tab:volumes}.

\subsection {Correlation Lengths}
\label{sec:xi}

\begin {figure}
\setlength\unitlength{1 in}
\vbox
   {%
   \begin {center}
   \begin {picture}(5,3.5)(0.1,0.3)
      \put(-0.2,0.3){
        \leavevmode
        
        \epsfbox [150 60 500 500] {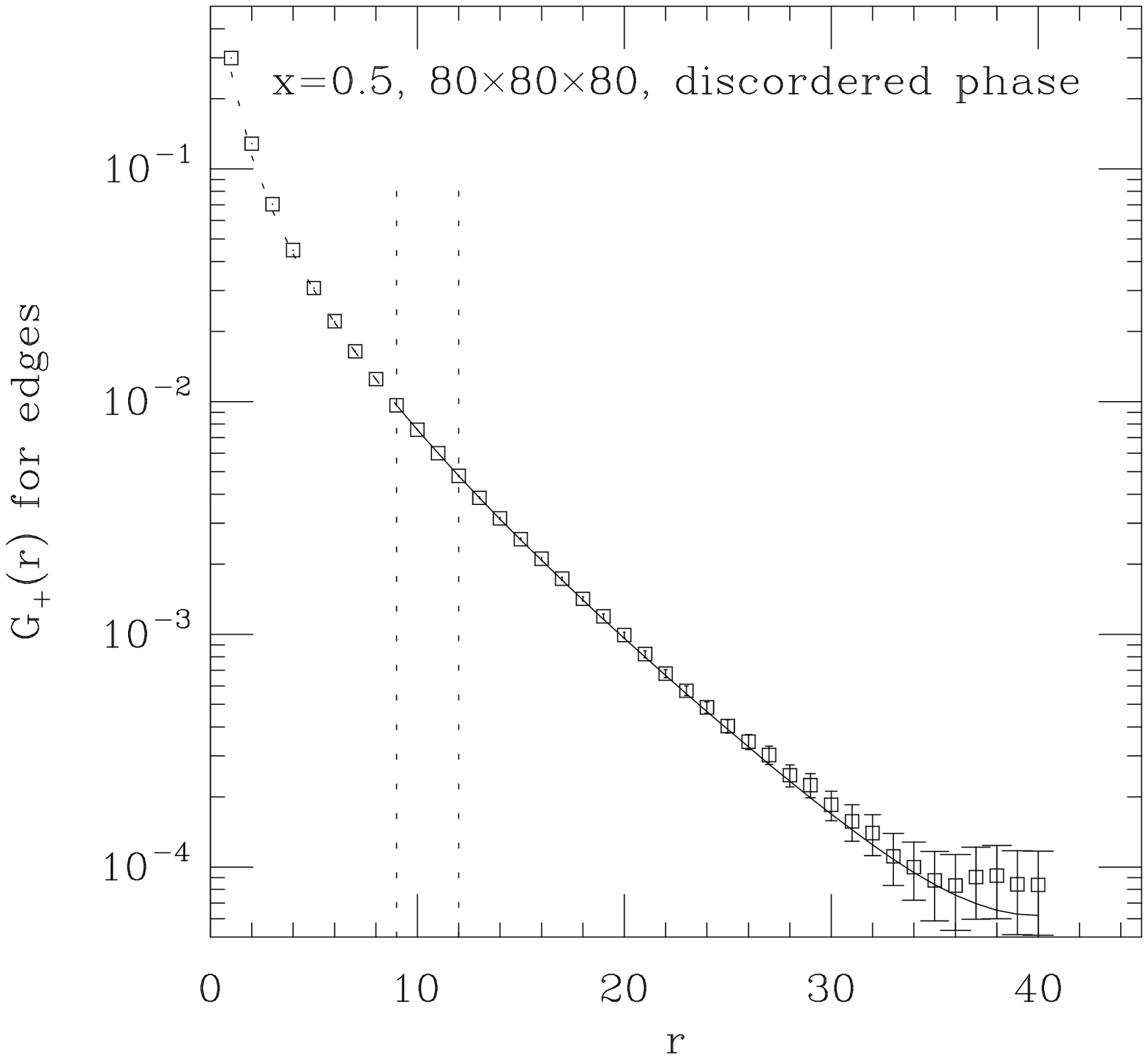}
      } 
      \put(3.2,0.3){
        \leavevmode
        
        \epsfbox [150 60 500 500] {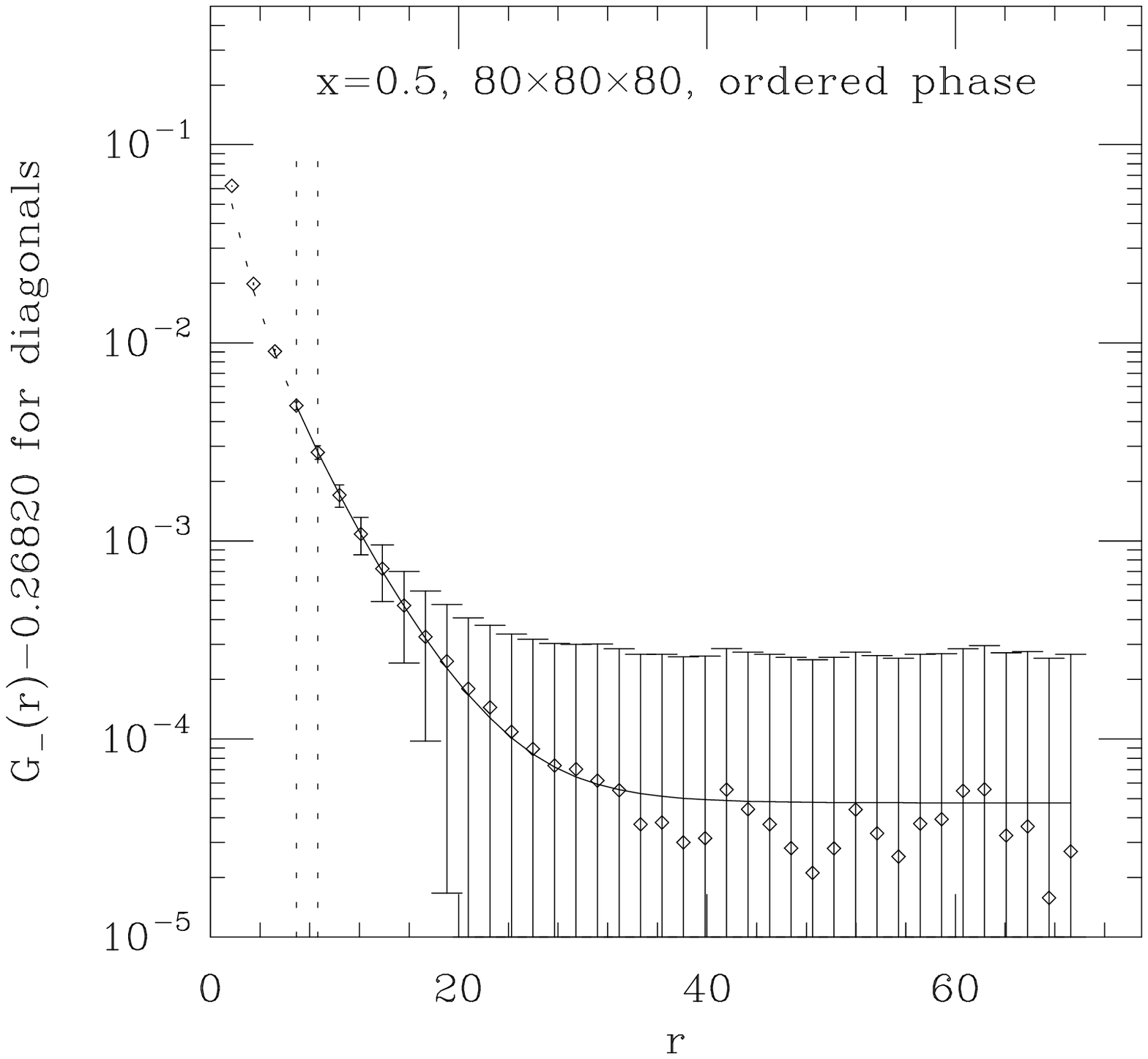}
      }
      \put(0,0.8){(a)}
      \put(3.4,0.8){(b)}
   \end {picture}
   \end {center}
   \caption
       {%
       Examples of fits to the correlation function $G(r)$ in (a) the
       disordered and (b) ordered phase.  The scatter of points around the
       fits is small because of correlations between the points.
       The dotted vertical lines indicate the first point used with
       the criteria
       $r > \xi$ (left line, and the value used for the fit)
       or $r > 1.5 \xi$ (right line).
       \label{fig:correlator}
       }%
   }%
\end {figure}

  To measure correlation lengths, we first measure the correlation
function
\begin {equation}
   G(r) = \langle {\bf S}(0) \cdot {\bf S}({\bf r}) \rangle \,,
\end {equation}
where, in practice, we average the right-hand size over all translations
and cubic rotations.  We measure this correlation for the two cases
of (1) displacements ${\bf r}$ that lie along edges of the lattice
({\it i.e.}\ in the direction of nearest neighbors), and (2) those that
lie along diagonals of the lattice.  In each case, we also measure the
full statistical error matrix of our measurements, which includes
correlated errors between different $r$.
(See sec.~\ref{sec:statistics} for our method of computing statistical
errors.)
Then we fit $G(r)$ to the form
\begin {equation}
   G(r) = a_+ \left[
             {1\over r} e^{-r/\xi_+} + (\hbox{images}) \right]
\end {equation}
for the disordered phase and
\begin {equation}
   G(r) = {a_-} \left[
              {1\over r} e^{-r/\xi_-} + (\hbox{images}) \right] + b_-
\label{eq:G disordered}
\end {equation}
for the ordered phase.
``Images'' denotes the similar exponential contributions from nearby
images of ${\bf r}$ due to the finite periodicity
of the lattice.
We first try fitting the above forms to all the data points in the
relevant direction (edge or diagonal).  Then we iteratively throw
away the smallest $r$ point from our data set until both
(1) the confidence level of the fit is at least 10\%,
and (2) $r > \xi$ for all the points being fit.
Fig.~\ref{fig:correlator} shows examples of fits to the correlation function.

\begin {figure}
\setlength\unitlength{1 in}
\vbox
   {%
   \begin {center}
   \begin {picture}(5,3.5)(0.1,0.3)
      \put(-0.2,0.3){
        \leavevmode
        
        \epsfbox [120 25 500 450] {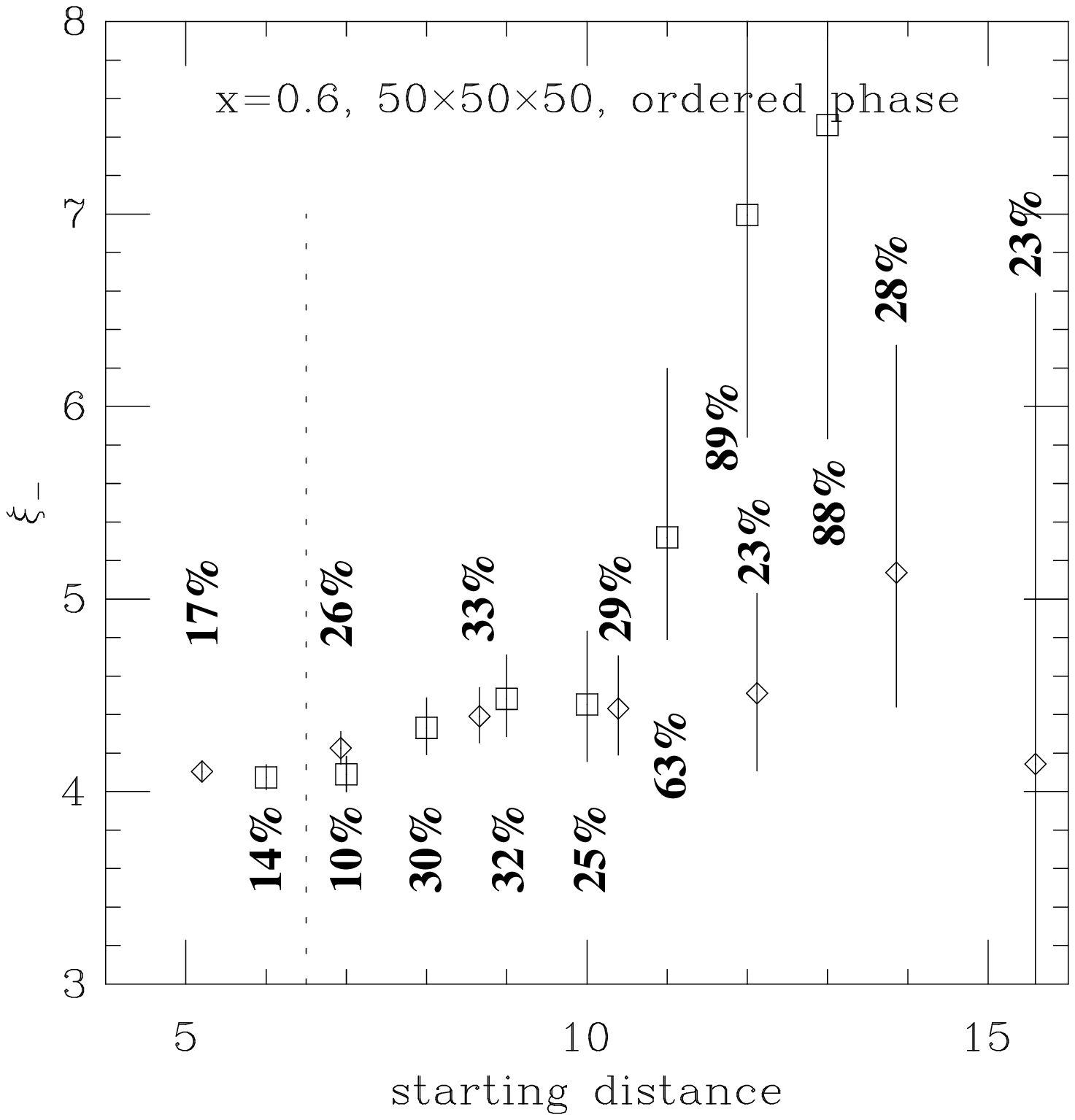}
      } 
      \put(3.2,0.3){
        \leavevmode
        
        \epsfbox [120 25 500 450] {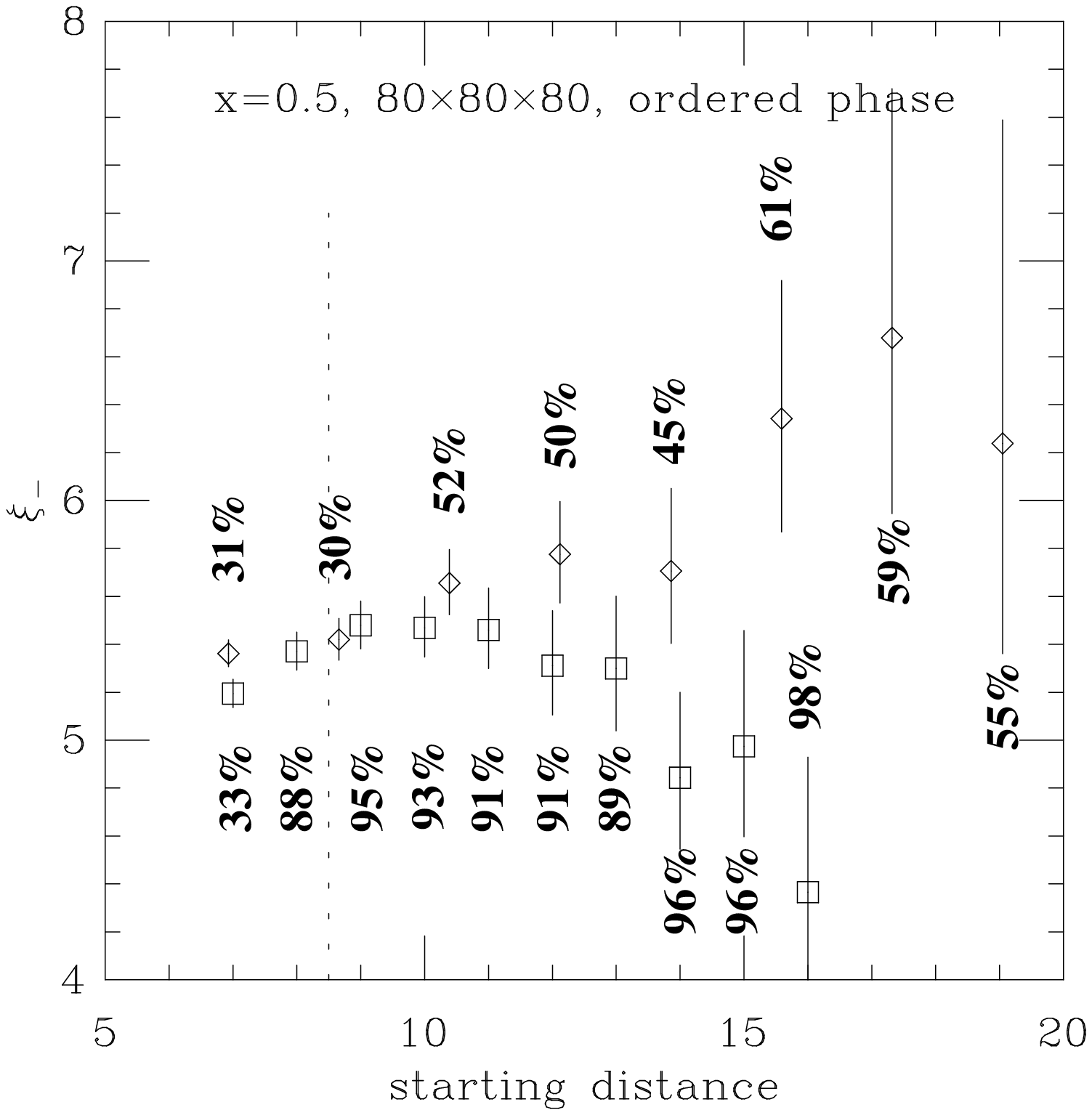}
      }
      \put(0.,2.5){(a)}
      \put(3.4,2.5){(b)}
   \end {picture}
   \end {center}
   \caption
       {%
         Examples of the dependence of our fits for $\xi$ on the
         minimum $r$ kept in the fit.  We show $\xi_-$ for particular
         runs at $\betat$ for (a) $x{=}0.6$ and (b) $x{=}0.5$.
         $\xi_-$ is measured both along edges (squares) and diagonals
         (diamonds).  The left-most point of each type is the point
         chosen for our final value, as described in the text.
         (The left-most point right of the dotted line is chosen
         if the criteria $r > 1.5\xi$ is used instead of $r>\xi$.)
         Confidence levels of the fit are given for each point.
         \label{fig:xi_vs_rmin}
       }%
   }%
\end {figure}

Fig.~\ref{fig:xi_vs_rmin}
shows the effect of continuing to throw out even more and more
short-distance points once our criteria are satisfied.
Focusing on
the fit to $\xi_-^\edge$,  there is
a clear drift of $\xi$ as more points are removed,
and then there is a plateau that is high compared to our nominal
value of $\xi$ and its error (the left-most data point  ).
One suspects that the value of $\xi$
at this plateau might be a better estimate than that from our procedure.
Unfortunately, we do not have a single, universal
criterion that would exactly agree with one's best subjective guess
of $\xi$ for every data set.
The drift in values suggests that the
systematic errors from fitting may be a bit larger than the
(statistical) errors we have assigned to $\xi$.
If one strengthened the requirement $r>\xi$ for points being fit
to $r>1.5\xi$, this drift would change our final interpolations for
the $x\to 0^+$ values of $\xi_+/\xi_-$ from (\ref{eq:xifit}) to
\begin {mathletters}%
\begin {equation}
   \lim_{x\to0^+} \> {\xi_+^\edge\over \xi_-^\edge} = 1.55(5) \,,
   \qquad
   \lim_{x\to0^+} \> {\xi_+^\diag\over \xi_-^\diag} = 1.58(6) \,,
\end {equation}%
for extrapolation vs. $\xi_+^{-\omega}$ and
\begin {equation}
   \lim_{x\to0^+} \> {\xi_+^\edge\over \xi_-^\edge} = 1.60(6) \,,
   \qquad
   \lim_{x\to0^+} \> {\xi_+^\diag\over \xi_-^\diag} = 1.64(9) \,,
\end {equation}%
\end {mathletters}%
for extrapolation vs. $\xi_-^{-\omega}$.  This is comparable to the
spread of values (\ref{eq:xifit}) and consistent with our final
result (\ref{eq:xiratio_result}).

The analysis of the correlation lengths for x=0.3 is slightly complicated.
Because of the large lattice volume required, it takes longer to generate
independent configurations than for other $x$.  At the same time, we can
measure the correlation function $G(r)$ at a larger number $N_r=60$ of values
of $r$.  If the number of independent configurations is large compared
to $N_r$, one can estimate the full covariance matrix
(see sec.~\ref{sec:statistics})
for all of these
measurements and use it to find the correlation length.
In our simulations,
however, the number of independent configurations $n$ generated for
$x=0.3$ (roughly 70 for the correlation function) is too small for this.%
\footnote{
  When $n \leq N_r$, for example, the measured covariance matrix will always be
  singular.
}
We have chosen to circumvent this issue by simply
throwing away many of our $r$ values when fitting the correlation function,
as we shall describe below.  As we shall see, the $x=0.3$ results do not
much affect our final results for $\xi_+/\xi_-$; so it is not necessary to
make a more sophisticated analysis.

\begin {figure}
\vbox
   {%
   \begin {center}
        \leavevmode
        
        \epsfbox [0 60 500 500] {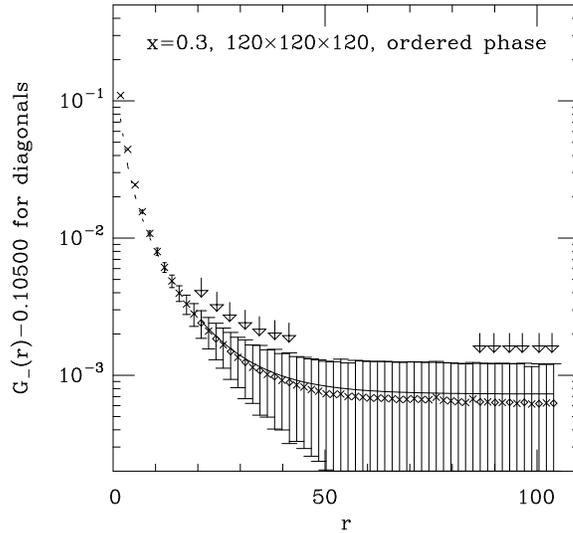}
   \end {center}
   \caption
       {%
       Correlation function $G(r)$ along diagonals for $x=0.3$
       in the ordered phase at the transition.
       The diamonds represent the points actually used for the fitting and
       are also marked by arrows.
       \label{fig:correlator0.3}
       }%
   }%
\end {figure}

Fig.~\ref{fig:correlator0.3}
shows the correlation function along diagonals for $x=0.3$
at the transition in the ordered phase.  Two regions of $r$ plausibly
contain the most important information for fitting the correlation
function: the points at the largest $r$, which determine $b_-$ in
(\ref{eq:G disordered}),
and the points from one to a few correlation lengths, which determine
$\xi_-$.  We have chosen to keep only the $r$ values marked by diamonds
in fig.~\ref{fig:correlator0.3}:
every other point for 6 points at the largest $r$, and
every other point for 7 points starting from $r\approx\xi$.  
We then have roughly five times as many independent measurements as 
$r$ points.
We use essentially the same prescription for the disordered 
phase and for the $\xi_\pm^\edge$.%
\footnote{
   Actually, our somewhat arbitrary criteria was to always keep $n/N_r$
   as close as possible to 5.  As a result, in some cases we took only 6
   instead of 7 points starting from $r\approx\xi$.
}
The results previously quoted in sec.~\ref{sec:analysis} were obtained
using this method.

We have checked the sensitivity of our results to the choice of
which points to keep.  If we instead simply take 12 to 13 evenly 
spaced points between $r = \xi$ and $r_{\rm max}$, 
we obtain $\xi_+/\xi_- = 1.42(7)$ for edges and 1.43(8) for diagonals
at $x=0.3$, as
compared to 1.45(7) and 1.41(8) listed in table~\ref{tab:ratios}.
Due to the large error in our x=0.3 results, this change of prescription has 
negligible effect on the extrapolated ratios of $\xi_+/\xi_-$.

\subsection {Finite volume effects}
\label{sec:volume}

Table~\ref{tab:volumes}
shows the largest lattice sizes we have used in the determination
of various quantities.  For all measurements, every lattice dimension
is at least five times the correlation length and is typically
ten times the correlation length.  We shall assess whether
finite volume errors are significant compared to our statistical errors.
Finite size effects should generally fall exponentially with increasing lattice
size.

\begin{table}
\def\cen#1{\multicolumn{1}{c}{$#1$}}
\def\cenb#1{\multicolumn{1}{c|}{$#1$}}
\def\qq{\multicolumn{1}{c}{''}}
\def\qqb{\multicolumn{1}{c|}{''}}
\def\ph{\phantom{1}}
\def\tim{{\times}}
\begin {center}
\tabcolsep=6pt

\begin {tabular}{|l|cll|c|}                                      \hline
$x$   &\cen{\>C_\pm,\xi_\pm,\chi_+}
                &\cen{\chi_-}   &\cenb{\betat}
 & \multicolumn{1}{c|}{conversion} \\ \hline
1.0   &  $40^3$ &  $160\tim40^2$   &  $160\tim40^2$
 & $40\simeq13.9\xi_+^\edge$  \\
0.8   &    ''   &       \qq        &      \qqb     
 & $40\simeq11.0\xi_+^\edge$  \\
0.6   &  $50^3$ &  $200\tim50^2$   &      \qqb
 & $50\simeq9.4\xi_+^\edge\ph$   \\
0.5   &  $80^3$ &  $240\tim60^2$   &  $240\tim60^2$
 & $80\simeq11.0\xi_+^\edge$  \\
0.3   & $120^3$ &  $240\tim120^2$  &  $480\tim120^2$
 & $120\simeq5.4\xi_+^\edge\ph$  \\ \hline
\end {tabular}
\end {center}
\caption
    {%
    \label {tab:volumes}
    Largest volumes used for various measurements.  The last column gives
    a conversion between a typical lattice dimension and the disordered
    phase correlation length $\xi_+^\edge$ (which is larger than $\xi_-$).
    }%
\end{table}

In many Monte Carlo applications,
the cleanest way to show that finite volume effects are negligible
for a given lattice volume is to repeat the simulation on smaller
and smaller volumes until the effects are clearly noticeable.
One then extrapolates the finite size corrections back to the original,
large volume.  Unfortunately, this procedure is problematical in
our case.  In smaller volumes, the system undergoes transitions
between the ordered and disordered phases.  One can still measure
quantities such as $C_\pm$ during a time period between transitions,
but, as the volume and that transition time decreases, the statistical
error of the measurement will increase.  This degradation of the
statistics for smaller volumes obscures the finite size effects
one would like to measure.

\begin {figure}
\setlength\unitlength{1 in}
\vbox
   {%
   \begin {center}
   \begin {picture}(5,7)(0.1,0.3)
      \put(-0.2,3.3){
        \leavevmode
        
        \epsfbox [100 60 500 500] {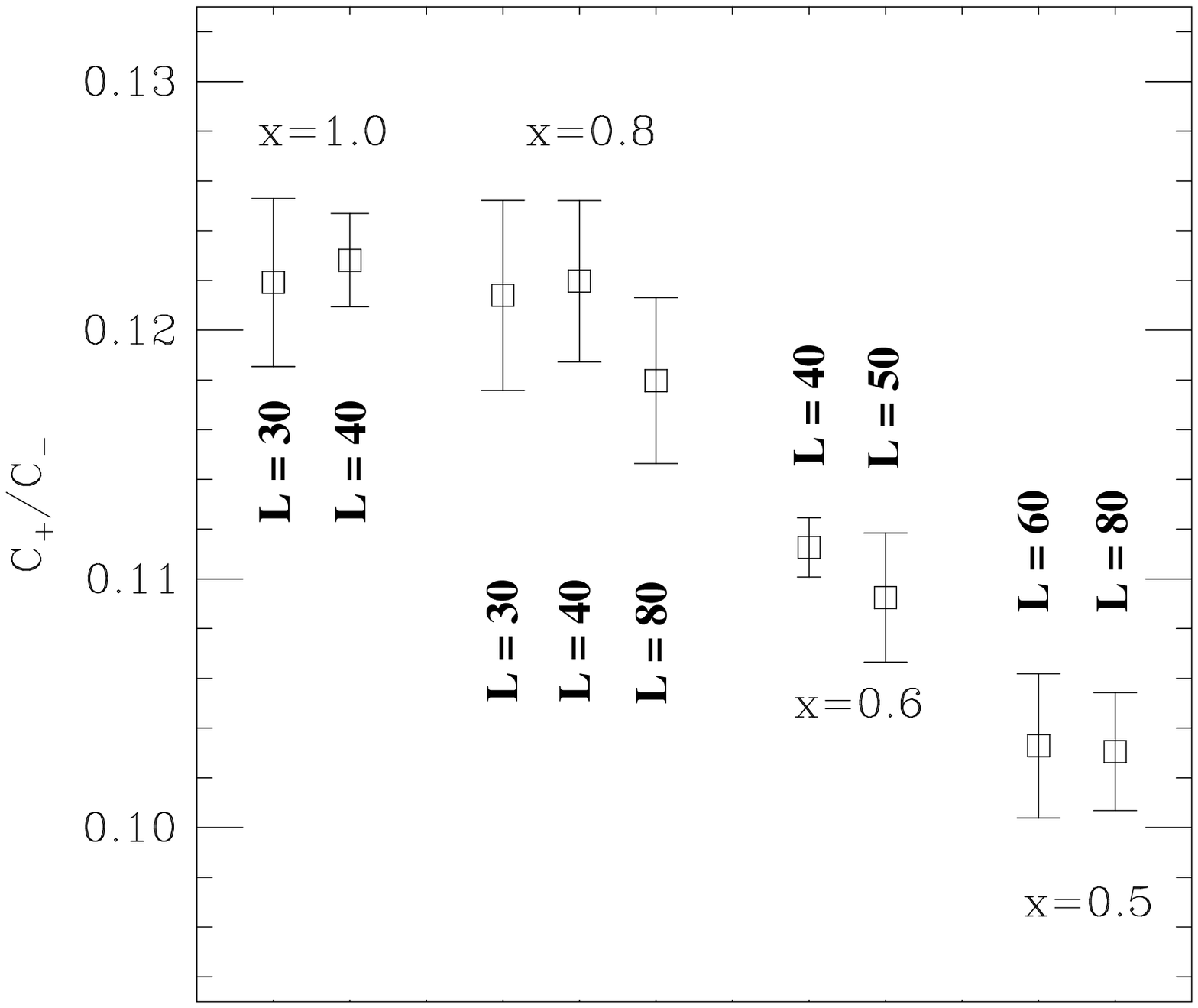}
      } 
      \put(3.2,3.3){
        \leavevmode
        
        \epsfbox [100 60 500 500] {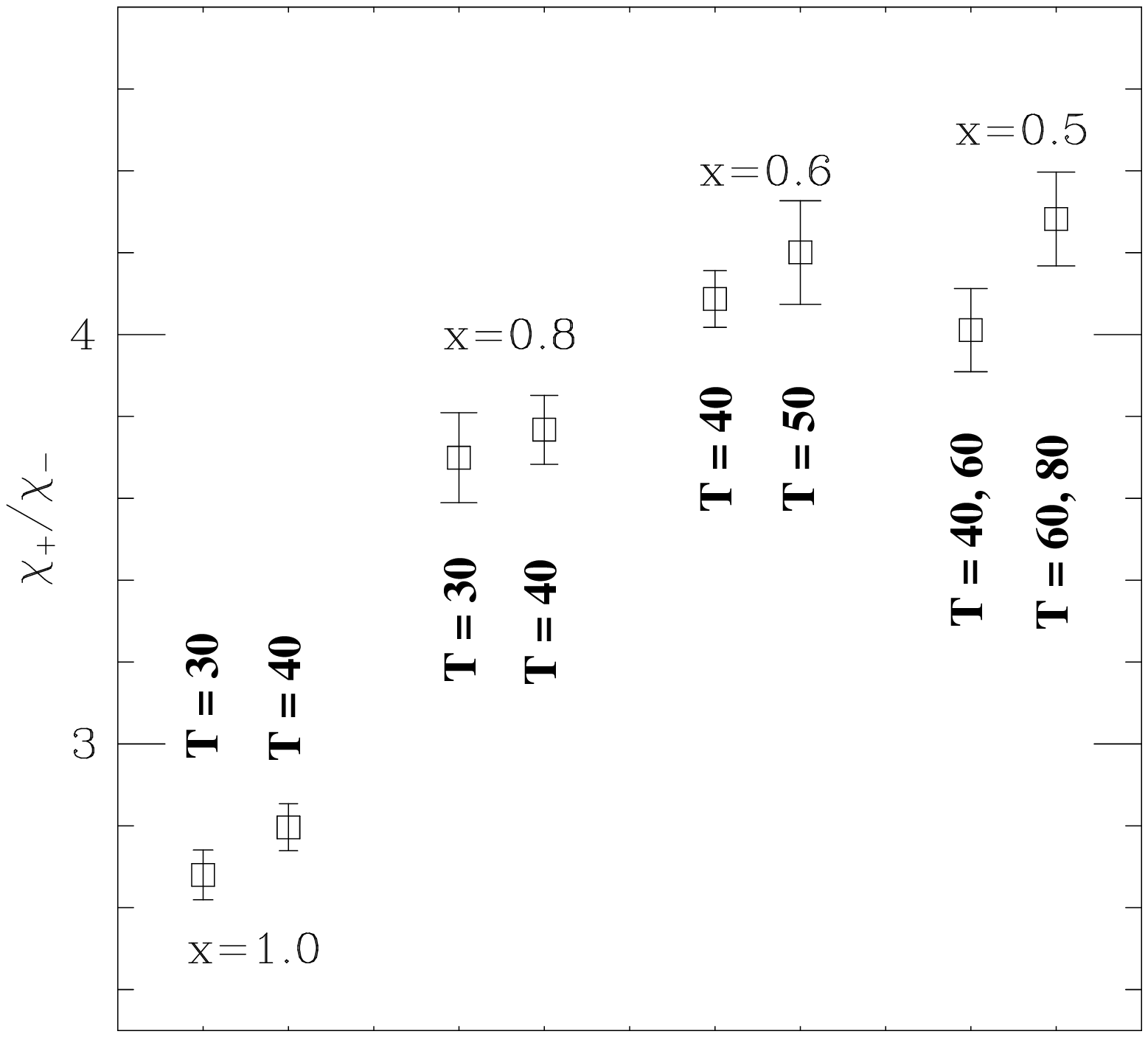}
      }
      \put(1.5,0.3){
        \leavevmode
        
        \epsfbox [100 60 500 500] {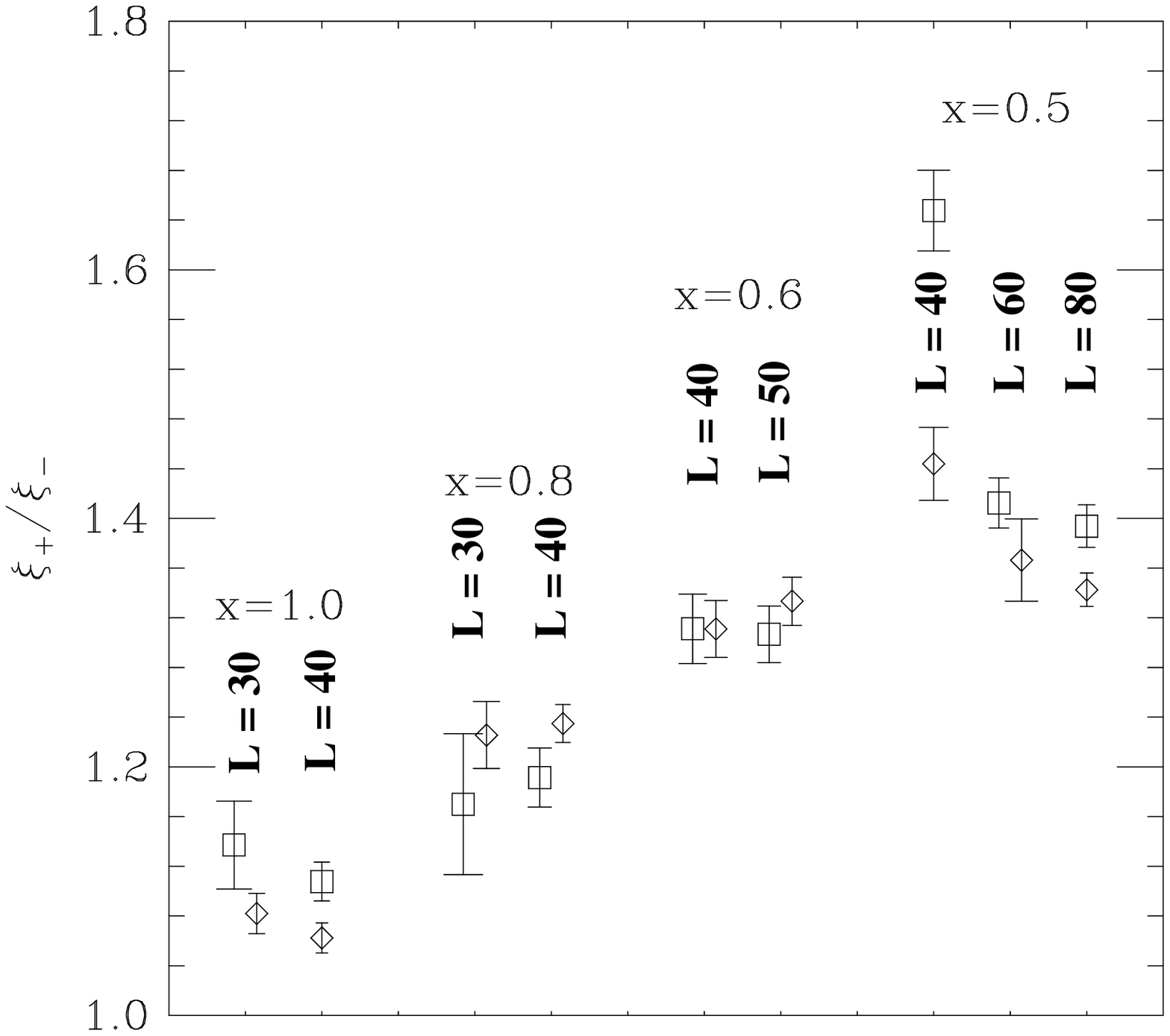}
      }
   \end {picture}
   \end {center}
   \caption
       {%
        \label {fig:ratios_vs_V}
           Checks of volume dependence of our ratios for all
           of our values of $x$.  $C_\pm$ and $\xi_\pm$ were measured on an
           $L^3$ lattice.  $\chi_+$ was measured on a $T^3$
           lattice, and $\chi_-$ was extrapolated on $L\times T^2$
           lattices.  Where there are two values of $T$ specified,
           the first is for $\chi_-$ and the second for $\chi_+$.
           For the $\xi$ ratio, the squares and diamonds are the results
           along edges and diagonals respectively.
           Error bars do not include the uncertainty in $\betat$.
       }%
   }%
\end {figure}

Instead, we content ourselves with simulating the system for a few
different ``large'' volumes and checking whether the discrepancies
seem consistent with statistical error.
Fig.~\ref{fig:ratios_vs_V} shows our checks.
The data for $C_+/C_-$ and $\xi_+/\xi_-$ look pretty good.
The data for $\chi_+/\chi_-$
suggests there might be a systematic effect making our larger
volume measurements slightly larger than our lower volume measurements.
In any case, we estimate that our finite size errors are no larger than
our statistical errors.

\begin {figure}
\vbox
    {%
    \begin {center}
        \leavevmode
        
        \epsfbox [150 60 500 500] {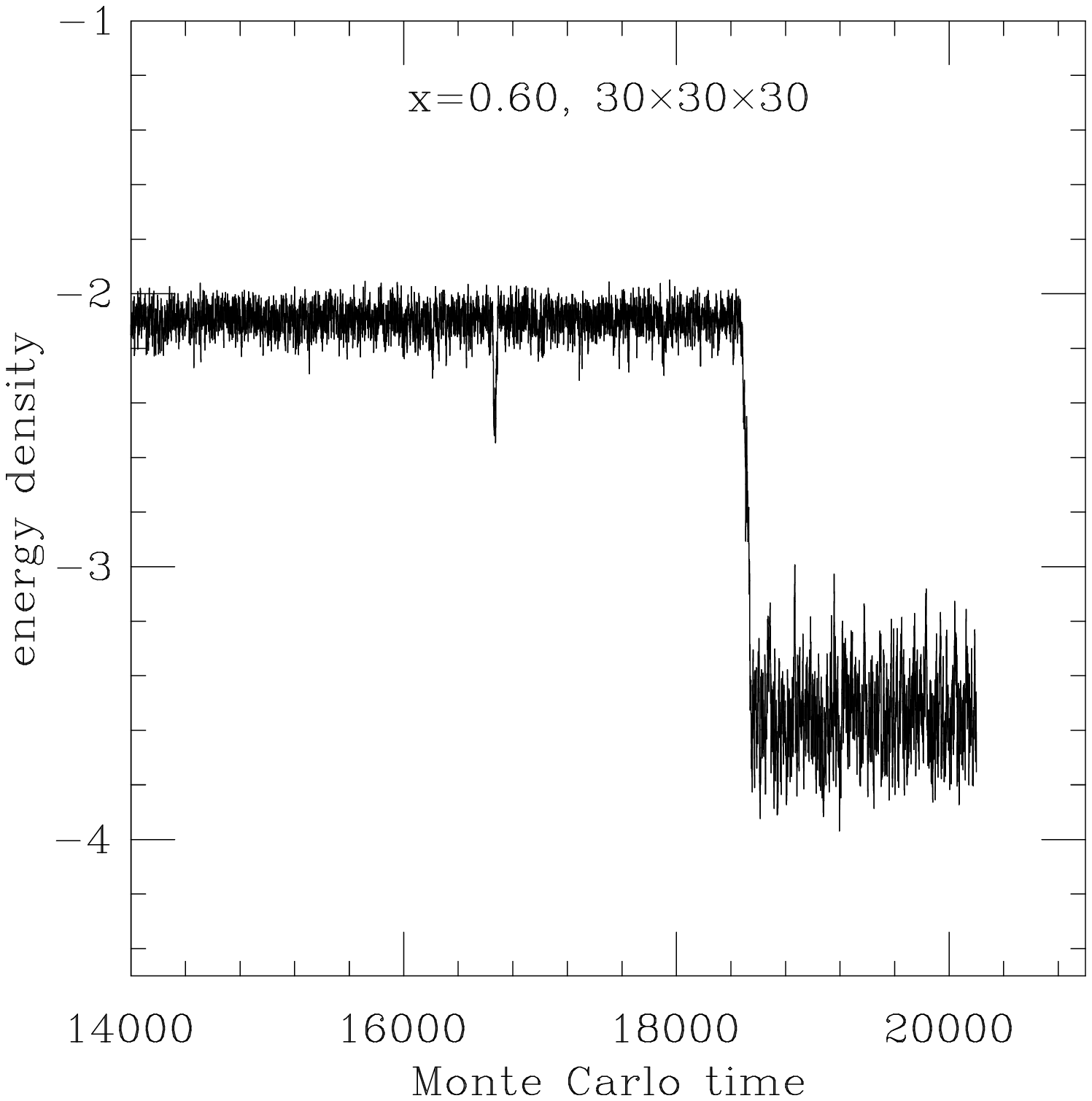}
    \end {center}
    \caption
        {%
        \label {fig:burp}
          The internal energy vs. Monte Carlo time for an $x{=}0.6$
          run on a somewhat smallish lattice ($30^3$).  This section of
          the run shows a transition between the phases and, preceding
          it, a spike corresponding to an unsuccessful transition attempt.
        }%
    }%
\end {figure}

We do not have data for $x{=}0.3$
but, based on the sizes of the correlation lengths, $x{=}0.3$ on
our $120^3$ lattice should be roughly comparable to $x{=}0.6$ on
a $30^3$ lattice.  A portion of a run on the latter is shown in
fig.~\ref{fig:burp}.  Not only do we see a transition between the
phases, but we also see a small spike corresponding to an
aborted transition attempt.  Whether or not such spikes appear
in one's data can have a significant effect on the extraction of
the specific heat.  For instance, if the run shown were cut off just
before the spike, one would find $C_+ = 2.36(4)$.  If the run were
cut-off just after the spike, one would find $C_+ = 2.51(10)$ and
suddenly have drastically revised the error estimate.
It is because of this sort of finite-size effect that we distinguish our
$x=0.3$ data as slightly less reliable.

In a few cases, we have checked the possibility of finite size errors
in our determination of $\betat$.  (Determining $\betat$ is quite time
consuming.) 
Table~\ref{tab:volumes betat} shows the results for
different choices of the transverse size $T$ of the $L\times T\times T$
lattices we use for determining the transition temperature.
The values are consistent with each other and, based on the transverse
sizes of our lattices in units of $\xi_+$
(as given by table~\ref{tab:volumes}), we believe that our
measurements for other $x$ should be reliable as well.

\begin{table}
\def\cena#1{\multicolumn{1}{c}{#1}}
\def\cenb#1{\multicolumn{1}{c|}{$#1$}}
\begin {center}
\tabcolsep=6pt

\begin {tabular}{|l|ll|}                    \hline
$x$   &  \cena{lattices}&  \cenb{\betat} \\ \hline
1.0   &  $L\times20^2$  &  0.157156(12)  \\
      &  $L\times40^2$  &  0.157154(4)   \\ \hline
0.5   &  $L\times40^2$  &  0.186750(4)   \\
      &  $L\times60^2$  &  0.186751(3)   \\ \hline
0.3   &  $L\times80^2$  &  0.200362(3)   \\
      &  $L\times120^2$ &  0.2003659(15) \\ \hline
\end {tabular}
\end {center}
\caption
    {%
    \label {tab:volumes betat}
      Dependence of the determination of $\betat$ on transverse
      lattice size for those $x$ where we measured it.
    }%
\end{table}


\subsection {Transition temperature uncertainties}
\label{sec:beta}

\subsubsection{Determining $\betat$}

\begin {figure}
\vbox
    {%
    \vspace* {-17pt}
    \begin {center}
        \leavevmode
        
        \epsfbox [150 250 500 550] {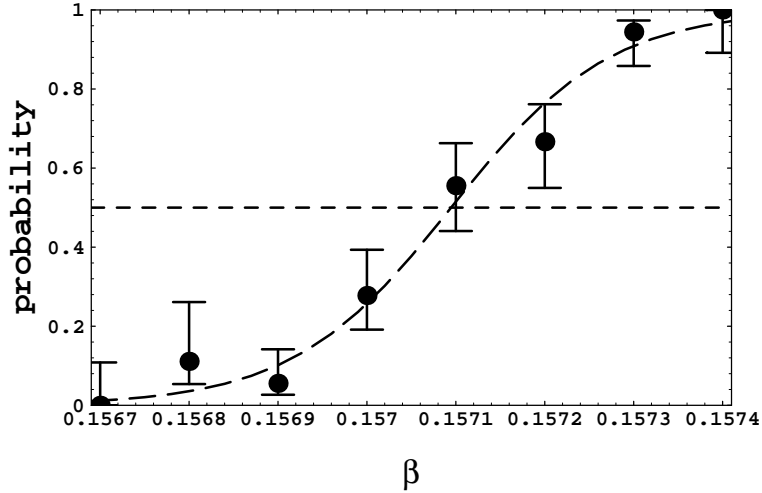}
    \end {center}
    \caption
        {%
        \label {fig:window}
           An example of the probability of ending in the ordered phase
           vs.\ $\beta$ when starting with mixed initial conditions.
           The data is for $x=1.0$ (the 4-state Potts model) on a
           $40\times20^2$ lattice.
           The fit (dashed line) to (\ref{eq:prob}) has a 75\% confidence
           level.  (This is typical of our fits to other data, at smaller
           $x$ or different lattice spacings, where the confidence levels
           of samples we checked ranged randomly from roughly 45\% to 85\%.
           At smaller $x$, we generally have somewhat fewer measurements and
           hence somewhat larger statistical uncertainty than shown above.)
        }%
    }%
\end {figure}

  As discussed in the introduction, our procedure for determining the
transition temperature is to make multiple runs starting from
mixed-phase initial conditions on long, asymmetric lattices
($L\times T\times T$) and to find the $\beta$ for which the
system is equally likely to end up in the ordered or disordered
phase.  A simple, biased random walk model of this process is presented
in appendix~\ref{app:gambler}, which predicts that the probability
$P$ of ending up in the ordered rather than disordered phase should
have $\beta$ dependence of the form:
\begin {equation}
   P \simeq {\textstyle{1\over2}} \left[
       1 + \tanh\left(\beta-\betat \over \Delta\beta\right) \right] \,,
\label{eq:prob}
\end {equation}
where $\betat$ and $\Delta\beta$ are not determined by the model and
$\Delta\beta$ may depend on the transverse lattice size
and on the Monte Carlo algorithm.
An example of our data, and the tanh curve that best fits it, is
shown in fig.~\ref{fig:window}.

At the beginning of each simulation, we obtain the initial condition
of fig.~\ref{fig:mixed-lattice} by initializing one half of the
lattice with $\beta{=}0$ initial conditions, one half with
$\beta{=}\infty$ initial conditions, and then evolving the two
halves independently for roughly 1000 sweeps (depending on $x$)
at the desired $\beta$.
Only then are the two halves allowed to interact and the interface
allowed to move, and the entire system is then evolved together.

\begin {figure}
\vbox
    {%
    \begin {center}
        \leavevmode
        
        \epsfbox [120 30 480 480] {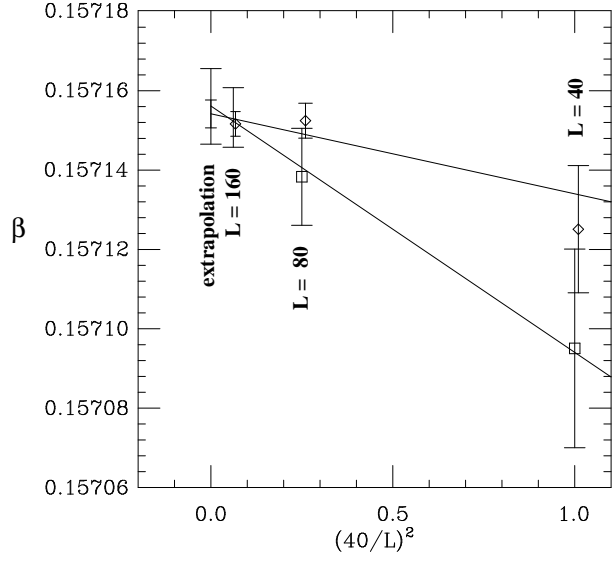}
    \end {center}
    \caption
        {%
        \label {fig:beta_vs_L}
          An example of the finite $L$ dependence of our determination
          of $\betat$.  The data is for $x{=}1.0$ on
          (squares) $L\times20^2$ and (diamonds) $L\times40^2$ lattices.
          (For $L{=}160$, the smaller error bar is the diamond point.)
          The lines are the best fits to the form $a + b L^{-2}$, and
          the extrapolations
          (see table~\protect\ref{tab:volumes betat}) are shown
          at $L{=}\infty$.
        }%
    }%
\end {figure}

Appendix~\ref{app:gambler} also discusses why the $\betat$ determined
by fitting (\ref{eq:prob}) is not necessarily correct for finite $L$,
and the model predicts the correction to the true $\betat$ scales
like $1/L^2$.
We therefore fit $\betat$ for a variety of $L$ and have extrapolated
to get the final values of table~\ref{tab:data}.  An example is
shown in fig.~\ref{fig:beta_vs_L}.  Data is shown for two different
transverse lattice sizes, which have different $1/L^2$ corrections
but extrapolate to consistent $L\to\infty$ limits.

\begin {figure}
\vbox
    {%
    \begin {center}
        \leavevmode
        
        \epsfbox [120 30 480 480] {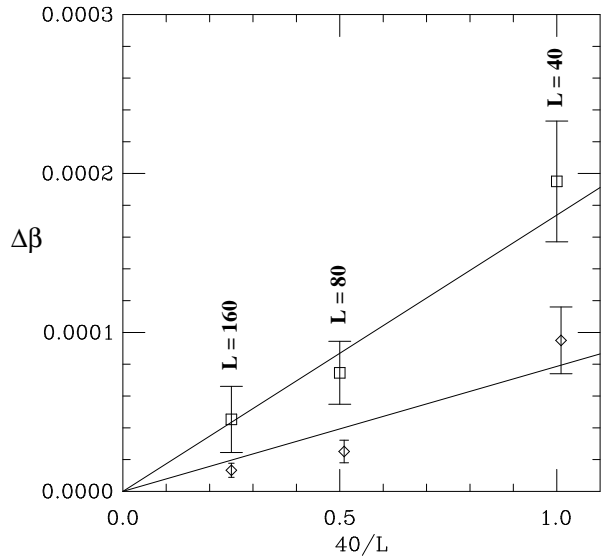}
    \end {center}
    \caption
        {%
        \label {fig:dbeta_vs_L}
          An example of the finite $L$ dependence of the width
          $\Delta\beta$ of our fits to tanh curves.  The data is
          for the same simulations as in
          fig.~\protect\ref{fig:beta_vs_L}.  The lines are the
          best fits to $a/L$.
        }%
    }%
\end {figure}

The simple model of Appendix~\ref{app:gambler} also predicts that the
width $\Delta\beta$ of the tanh curves should scale like $1/L$.  Though
this is not directly relevant to our determination of $\betat$, it is
worth checking.  Fig.~\ref{fig:dbeta_vs_L} shows a fit of this behavior
to the data corresponding to fig.~\ref{fig:beta_vs_L}.  The fit is
not very good for the $L\times40^2$ data, with 2\% confidence level,
and this mediocrity is typical
of our data at other values of $x$.  In contrast, our fits for $\betat$
typically work fairly well.
We suspect that the failure of our model for $\Delta\beta$ may be due
to the non-local nature of the cluster algorithm.

\subsubsection{Effects on other measurements}

Given our estimates of the uncertainty in the transition temperature
(see Table~\ref{tab:data}), we now need to determine the resulting
uncertainties in our measurements of $C$, $\xi$, and $\chi$.
For $C$ and $\chi$, we do this by measuring
\begin {eqnarray}
   \partial_\beta(\beta^{-2}C)
   &=& -N^2 \langle (\epsilon - \bar\epsilon)^3 \rangle
   \,,
\\
   \partial_\beta\chi
   &=& -N \left[ \langle \chi \eps \rangle
                 - \langle\chi\rangle\langle\eps\rangle \right] \,,
\end {eqnarray}
where $N$ is the lattice volume, $\epsilon = E/N$ is the energy density,
and $\bar\epsilon \equiv \langle\epsilon\rangle$.  We make our measurements
at the central value of $\betat$.
Results are given in table~\ref{tab:derivs}.
Our determination of these derivatives at small $x$
is not particularly good, but it only needs to be good enough to estimate
our error due to the uncertainty in $\betat$.
For the correlation length, we similarly compute
the $\beta$ derivatives of the correlation function and of our error
matrix.  The separate sizes of statistical errors and
$\betat$ uncertainty errors in our final results is given in
table~\ref{tab:errors}.

\begin{table}
\def\cen#1{\multicolumn{1}{c}{$#1$}}
\def\cenb#1{\multicolumn{1}{c|}{$#1$}}
\def\e#1{$\times10^{#1}$}
\begin {center}
\tabcolsep=6pt

\begin {tabular}{|l|cccc|}                                           \hline
$x$   &\cen{\partial_\beta(\beta^{-2}C_+)}
   &\cen{\partial_\beta(\beta^{-2}C_-)}
   &\cen{\partial_\beta\chi_+}
   &\cenb{\partial_\beta\chi_-} \\ \hline
1     & 2.14(17)\e4  & -4.0(6)\e5   & 1.29(6)\e4  & -1.39(20)\e4 \\
0.8   & 2.2(3)\e4    & -6.5(1.2)\e5 & 2.55(18)\e4 & -1.82(17)\e4 \\
0.6   & 4.0(5)\e4    & -9.4(1.9)\e5 & 8.7(9)\e4   & -6.2(5)\e4   \\
0.5   & 6.7(8)\e4    & -1.5(4)\e6   & 2.42(19)\e5 & -1.99(22)\e5 \\
0.3   & 4.9(1.2)\e5  & -2.3(1.0)\e7 & 1.30(24)\e7 & -1.21(23)\e7 \\ \hline
\end {tabular}
\end {center}
\caption
    {%
    \label {tab:derivs}
    Summary of $\beta$ derivatives of $C_\pm$ and $\chi_\pm$ at the
    central value of $\betat$.
    }%
\end{table}

\begin{table}
\def\cen#1{\multicolumn{1}{c}{$#1$}}
\def\cenb#1{\multicolumn{1}{c|}{$#1$}}
\def\e#1{$\times10^{#1}$}
\begin {center}
\tabcolsep=6pt

\begin {tabular}{|l|cccc|}                                           \hline
      & \multicolumn{4}{|c|}{$\delta\betat$ error / stat.\ error} \\
\cline{2-5}
$x$   & \multicolumn{1}{|c}{$C_+/C_-$}
      & \multicolumn{2}{c}{$\xi_+/\xi_-$}
      & \multicolumn{1}{c|}{$\chi_+/\chi_-$} \\
      &&\multicolumn{1}{c}{edge}&\multicolumn{1}{c}{diag.}& \\ \hline
1     &  0.2   &  0.2  &  0.3 & 0.3  \\
0.8   &  0.2   &  0.1  &  0.2 & 0.2  \\
0.6   &  0.2   &  0.3  &  0.3 & 0.3  \\
0.5   &  0.5   &  0.7  &  0.8 & 0.8  \\
0.3   &  0.4   &  0.5  &  0.4 & 0.4  \\ \hline
\end {tabular}
\end {center}
\caption
    {%
      \label {tab:errors}
      Relative size of error due to uncertainty in $\betat$ compared to the
      statistical error for the ratios whose total errors are given in
      table~\protect\ref{tab:ratios}.  The total error is the
      $\betat$ uncertainty added in quadrature with the statistical
      errors.  (For $\chi_-$, ``statistical error'' includes the
      extrapolation $L\to\infty$.)
    }%
\end{table}


\subsection {Statistical errors and algorithms}
\label{sec:statistics}

Statistical errors were determined by first computing the decorrelation
time $\taud$ relevant for each quantity.
(For example, for $C$, we use the decorrelation time in the energy.)
By decorrelation time, we mean a measure of the Monte Carlo time over
which configurations {\it within a given phase} become
statistically uncorrelated.  Our decorrelation time does not measure
the mixing time between phases, since we run in large enough volumes
that there are no transitions between phases in our simulations.

Consider first estimating the error of quantities which are determined
by simple ensemble averages of some quantity $A$ ({\it e.g.}
the susceptibilities $\chi(p)$ or the energy density $\epsilon$).
We then use the integrated decorrelation time defined by \cite{taud}
\begin {equation}
   \tau_{int} = {1\over2} \sum_{t=-\infty}^{+\infty} C(A;t) \,.
\label{eq:taud}
\end {equation}
$C(A;t)$ is the auto-correlation function, estimated for a sample of
$n$ measurements as \cite{timeseries}
\begin {equation}
   C(A;t) = {
     {\displaystyle
       {1\over n-t} \sum_{i=1}^{n-t} (A_i - \overline{A}) 
                                     (A_{i+t} - \overline{A})
     }
     \over
     \sigma_{A}^2
   } ,
\end {equation}
where $\overline{A}$ is the sample average of $A$ and
\begin {equation}
   \sigma_{A}^2 = {1\over n} \sum_{i=1}^n (A_i-\overline{A})^2
   .
\end {equation}
The error in $\overline{A}$ is then estimated as
\begin {equation}
   \hbox{Err}(\overline{A}) = {\sigma_A \sqrt{2 \tau_{int} \over n}}
\end {equation}

In practice, the sum in the definition (\ref{eq:taud}) must be cut off,
because the statistical error in C(A;t) itself becomes large for large
$t$.  Our criteria is to cut off the sum when $C(A;t)$ drops below
0.05.  The value 0.05 was chosen to give reasonable agreement with
the binning method described next.  Reducing 0.05 to 0.01 would not
appreciably change our results.

\begin {figure}
\vbox
   {%
   \begin {center}
        \leavevmode
        
        \epsfbox [150 60 500 500] {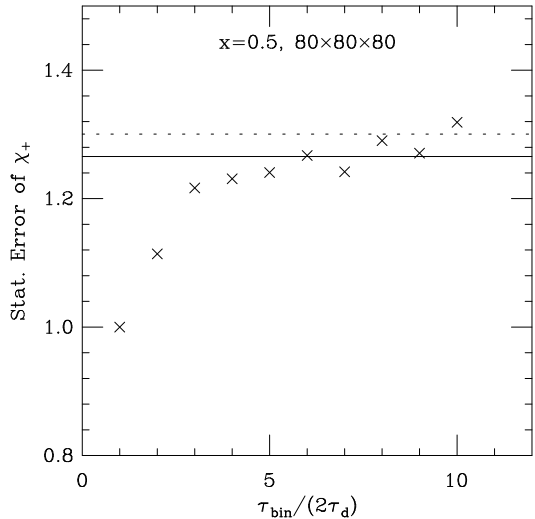}
   \end {center}
   \caption
       {%
       Example of statistical error estimates vs.\ bin size.
       The data is from measurements of $\chi_+$ at $x=0.5$.
       The vertical axis has been arbitrarily normalized to be
       1.0 for $\tau_{bin}=2\tau_{d}$.
       The solid horizontal line is the independent error estimate
       using the integrated decorrelation time.  The dotted line is
       what the same estimate would be if we had
       cut off the sum (\ref{eq:taud}) when $C(A;t)$ dropped below 0.01 rather
       than 0.05.
       \label{fig:binning_chi+}
       }%
   }%
\end {figure}

Alternatively, one can estimate errors by binning the sequence of
measurements $A_i$ ($i=1,n$) into bins of some size $\tau_\bin$ and averaging
the data over each bin to obtain a new sequence $\bar A_k$ ($k=1,N_\bin$),
where $N_\bin = n/\tau_\bin$.  The error is then estimated by the
variance of this new sequence:
\begin {equation}
   [\hbox{Err}(\overline{A})]^2 = {1\over(N_\bin - 1)}
     [\langle \bar A^2 \rangle - \langle \bar A\rangle^2 ] \,.
\end{equation}
Fig.~\ref{fig:binning_chi+}
shows a typical example of the result vs.\ bin size.
The error stabilizes at large bin sizes, as it should, and roughly 
agrees with the integrated decorrelation time method, which is also
shown in the figure.

\begin {figure}
\setlength\unitlength{1 in}
\vbox
   {%
   \begin {center}
   \begin {picture}(5,3.5)(0.1,0.3)
      \put(-0.2,0.3){
        \leavevmode
        
        \epsfbox [120 75 500 450] {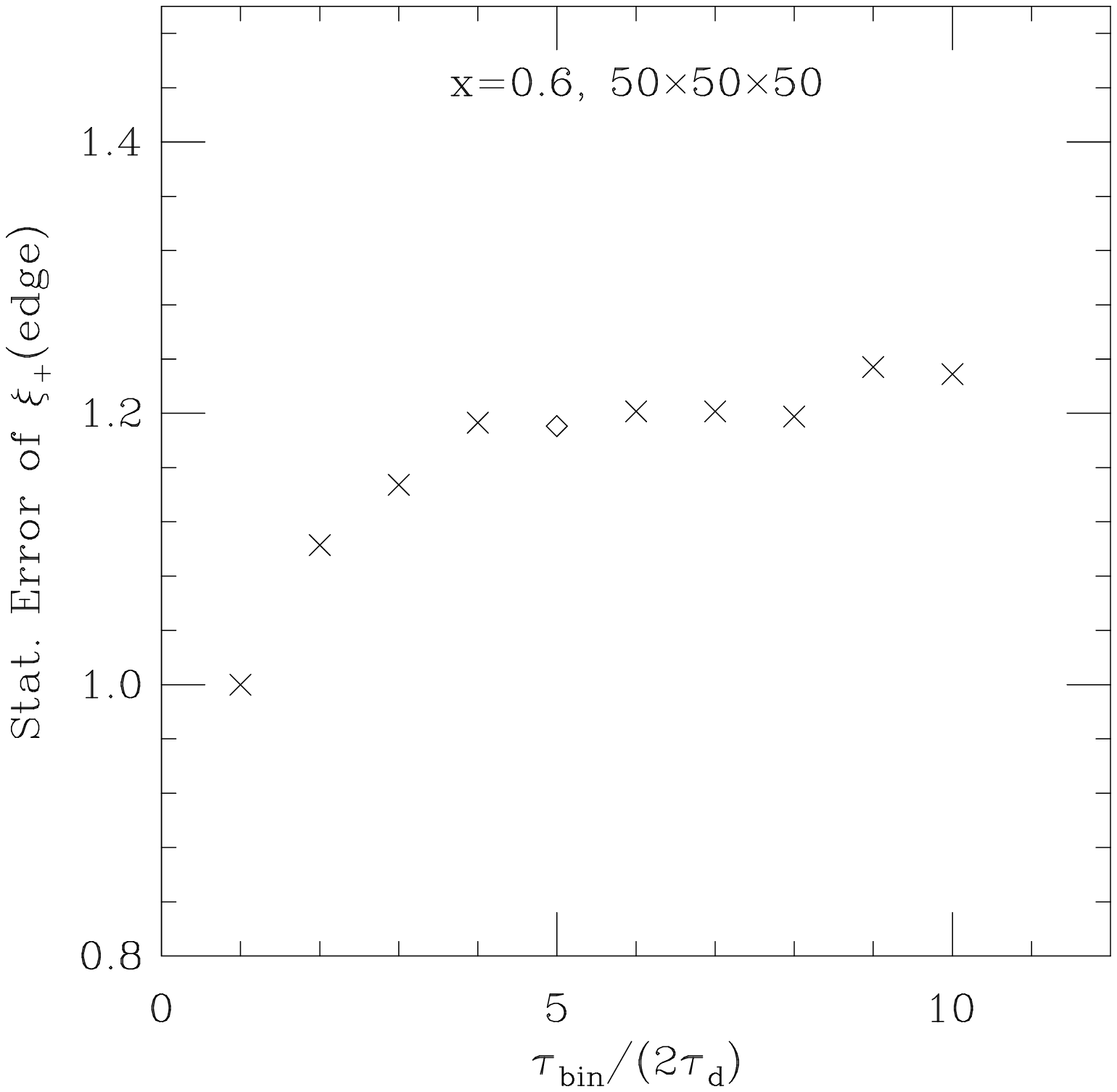}
      } 
      \put(3.2,0.3){
        \leavevmode
        
        \epsfbox [120 75 500 450] {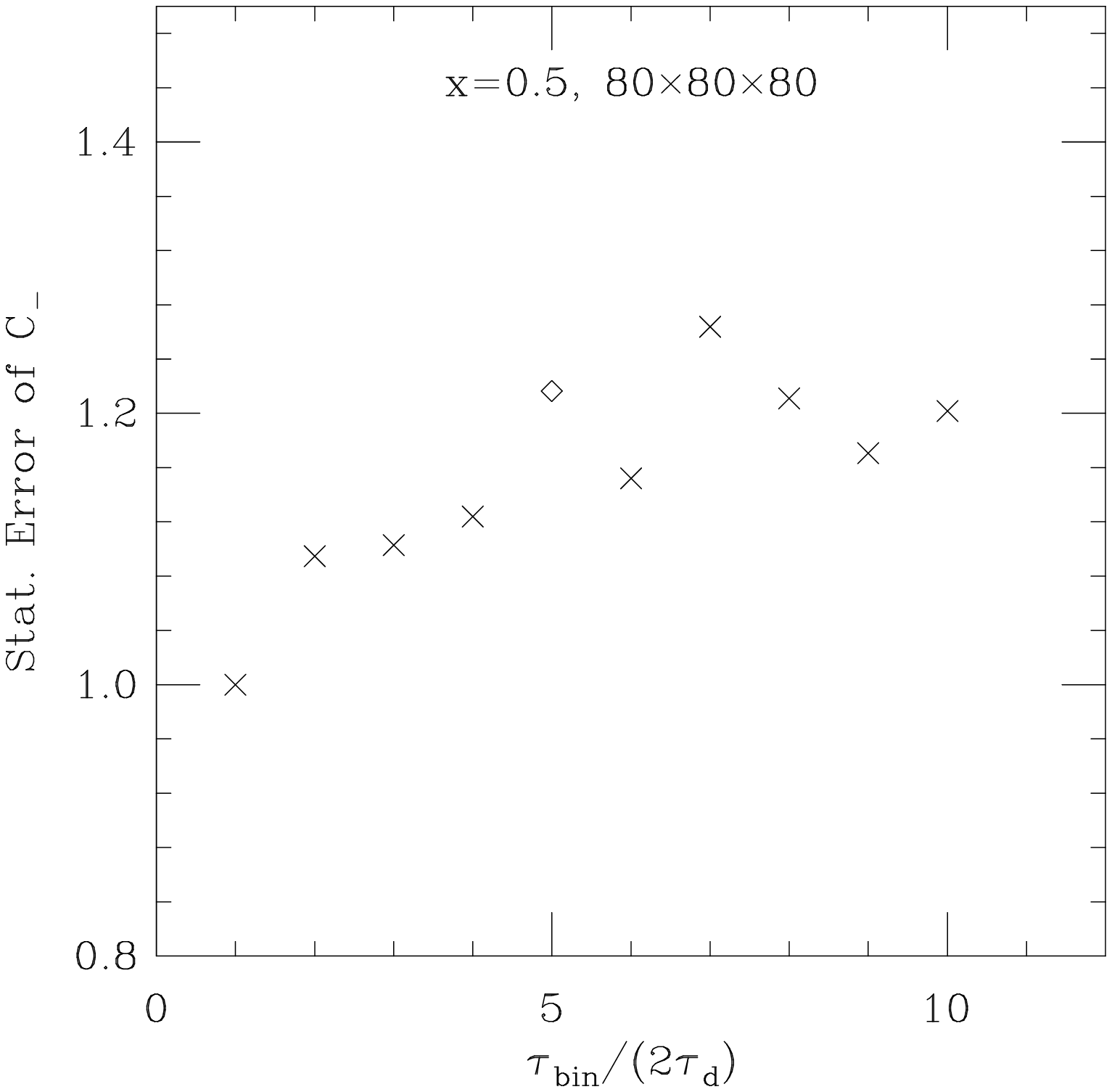}
      }
      \put(0.1,2.6){(a)}
      \put(3.5,2.6){(b)}
   \end {picture}
   \end {center}
   \caption
       {%
         Examples of the dependence of statistical error on bin size
         for (a) $\xi_+^\edge$, and (b) $C_-$.
         The vertical axis has been arbitrarily normalized to be
         1.0 for $\tau_{bin}=2\tau_{d}$.
         The diamond marks the actual bin size used according to our
         criteria.
         \label{fig:binning}
       }%
   }%
\end {figure}

To calculate correlation lengths, we need the full correlated error
matrix (the covariance matrix) of the correlation functions $G(r)$.
We compute this using the binning method.
Binning (with fixed bin size) has
the advantage that the resulting covariance matrix is positive definite.
The covariance matrix is given by
\begin {equation}
   \sigma_{ij} =
   {1\over(\Nbin-1)} [\langle\bar {\cal G}(r_i)
       \bar {\cal G}(r_j)\rangle
        - \langle\bar {\cal G}(r_i) \rangle
        \langle \bar {\cal G}(r_j) \rangle]
   \,,
\label{eq:G corr}
\end {equation}
where ${\cal G}(r) = {\bf S}({\bf r}) \cdot {\bf S}(0)$ 
averaged over translations and cubic rotations.
In order to have a simple universal criteria for what
size of bin to use, we have surveyed a variety of examples and found
that a bin size of $10 \taud$ works well, where $\taud$ is the maximum
(over the range of $r$ used for the fit) of the integrated
decorrelation time for the $G(r)$.
Fig.~\ref{fig:binning}a
shows a typical example of the dependence of the statistical error of
correlation length on bin size, and our particular choice of bin size 
can be seen to be adequately large.

The error estimate for the correlation lengths are determined by
a standard chi-square analysis.

One case that requires a slightly different approach is the
specific heat, which is measured from the variance of the
energy density.  In this case, we bin the energy density as above
but don't average it over each bin.  Then we apply the jackknife
procedure on the bins: we compute the specific heat after
throwing away the $\tau_\bin$ energy density measurements corresponding to
one of the bins.  The $\Nbin$ possibilities for which bin to have thrown
out gives us $\Nbin$ separate measurements of the specific heat.
The variance of these measurements is used for the error estimate on
the specific heat.
By surveying the dependence on bin size for all are data, we have found
that a good choice is $\tau_\bin = 10\taud$ where $\taud$ is the integrated
decorrelation time for the energy.  An example of the dependence on bin
size is shown in Fig.~\ref{fig:binning}b.

A similar approach is used for $\chi_-$ and all $\beta$ derivatives
({\it e.g.}\ $\partial_\beta\chi_\pm$).

We have checked that running simulations with a pure heatbath algorithm,
or with a pure cluster update algorithm, give results that are statistically
consistent with each other and with our interlacing of the two algorithms.%
\footnote{
  In fact, such checks led us to discover for ourselves the problems
  of naively using
  simple random-number generators.  We had initially used a 32-bit
  congruence algorithm with a period of $2^{32}$.  This period is too small
  for the length of some of our simulations, manifesting in inconsistency
  between the heatbath and cluster algorithms and producing correlations
  which could clearly be seen in the very-long-time tail of the energy
  auto-correlation function.
  We had to switch to a generator with period $2^{64}$.
}


\bigskip

This work was supported by the U.S. Department of Energy,
grants DE-FG06-91ER40614 and DE-FG03-96ER40956.
We thank Joseph Rudnick, Michael Fisher, Joan Adler, and David Wright
for useful conversations,
Peter Ungar for explaining the solution to the Gambler's Ruin problem,
Marcel Den Nijs for pointing out that the cross-over
exponents are determined by well-known Ising model critical exponents,
and, most of all, Steve Sharpe and Larry Yaffe, who gave too many valuable
suggestions to enumerate.

\appendix

\section{Cluster Update Algorithm}
\label{app:cluster}

The two of us have
used different versions of a cluster update algorithm,
one which grows and flips a single cluster at a time, and another which
grows simultaneous clusters across the entire lattice.  In any case, they
are simple generalizations of the algorithms
\cite{swendsen&wang,wolff} used in other spin
systems.  Write the Hamiltonian as
\begin {equation}
   \beta H = \sum_{\langle i j\rangle} h({\bf S}_i, {\bf S}_j)
\end {equation}
where $h({\bf S}_i,{\bf S}_j)$ is the nearest-neighbor interaction.

One step of the first version of the algorithm can be summarized as follows.
(a) Randomly choose one of the
five order-2 elements $R$ ($R^2=1$) of the internal symmetry group $D_4$.
(b) Randomly choose a lattice site $x$ as the first point to include
in the cluster $c$, and mark the site.
(c) One at a time, visit all new links $\langle x y \rangle$
connecting $x \in c$ to its nearest neighbors $y$.  For each link
visited, check if $y$ is already in $c$, and if not then adjoin it to $c$
with probability $P({\bf S}_x,{\bf S}_y)$, where
\begin {equation}
   P({\bf S}_x,{\bf S}_y) = 1 - \exp\left\{ \min[0, h({\bf S}_x, {\bf S}_y)
                              - h(R\,{\bf S}_x, {\bf S}_y) ] \right\} \,.
\label{eq:flip-prob}
\end {equation}
A newly included $y$ should be marked.
(d) Repeat step (c) until no new sites are added to the cluster.
(e) Flip all the spins in the resulting cluster:
${\bf S}\to R\,{\bf S}$.

Randomness of the choices in steps (a) and (b) is inessential:
one only needs to vary the choices enough to give reasonably efficient
ergodicity.
If (a) is restricted to a random choice between just the two symmetries
corresponding to $s\to-s$ and $t\to-t$ respectively, then this algorithm
is equivalent to the Ashkin-Teller cluster algorithm described in
ref.~\cite{wiseman&domany}.  (See also ref.~\cite{salas&sokal}.)

One step of the second version of the algorithm is as follows.
(a) Randomly choose an $R$ as above.
(b) Visit every link $\langle x y \rangle$ in the lattice and
mark it with probability $P({\bf S}_x,{\bf S}_y)$
given by (\ref{eq:flip-prob}).  (c) Identify all the disconnected
clusters of sites connected by marked links, and flip each such cluster
with probability 1/2.


\section {The Gambler's Ruin problem}
\label{app:gambler}

In section~\ref{sec:method}, we discussed how we determine the transition
temperature by splitting an asymmetric $L\times T\times T$ lattice in half
along the longitudinal direction $L$ and starting the two halves in the
ordered and disordered phases respectively.
In this appendix, we explain
a simple model for the probability of the system evolving into one phase
over the other.

At any time, let $z_1$ and $z_2$ be the (transversely averaged) locations
of the two domain walls, and let $z$ be the separation between them as
measured through the disordered phase (or $L-z$ as measured through the
ordered phase).  When $\beta{=}\betat$, $z$ should random walk in
Monte Carlo time.  When $\beta$ deviates slightly from $\betat$, there
will be a slight bias in this random walk proportional to
$\beta{-}\betat$.
If we model this biased random walk as taking fixed steps in $z$
with probability
\begin {eqnarray}
   {\rm prob}(z\to z{+}1) &\equiv& p = {\textstyle{1\over2}} (1 + \eps) \,,
\\
   {\rm prob}(z\to z{-}1) &\equiv& q = {\textstyle{1\over2}} (1 - \eps) \,,
\end{eqnarray}
where $\eps \propto \beta{-}\betat$ is small, then we have a special case
of the Gambler's Ruin problem.  The Gambler's Ruin problem is that
you start with $z$ dollars in your pocket and you play some casino game
over and over again until you either go broke or accumulate your goal of
$L$ dollars.  What is the probability you can afford the taxi home?

To solve it, let $P(z)$ be the probability of winning if you start at
$z$.  Then, by considering one step, one obtains the difference equation
\begin {equation}
   P(z) = p P(z+1) + q P(z-1) \,,
\end {equation}
and the boundary conditions are
\begin {equation}
   P(0) = 0 \,, \qquad P(L) = 1 \,.
\end {equation}
The solution is
\begin {equation}
   P(z) = { 1 - \left(q\over p\right)^z \over
            1 - \left(q\over p\right)^L } \,.
\end {equation}
In our case (small $\eps$ and large $L$), this becomes
\begin {equation}
   P(z) \simeq { 1 - e^{-2 \eps z} \over 1 - e^{-2 \eps L} } \,.
\label{eq:Pgeneral}
\end {equation}
Putting in our initial condition $z=L/2$, we find that the curve in
fig.~\ref{fig:ruin}a is a tanh curve:
\begin {equation}
   P(L/2) \simeq {\textstyle{1\over2}}
      [ 1 + \tanh\left(\textstyle{1\over2}\eps L\right) ] \,.
\label{eq:Pfinal}
\end {equation}
We don't know {\it a priori} the proportionality constant between
our parameter $\eps$ and $\beta-\betat$, but we can parametrize the
curve as
\begin {equation}
   P \simeq {\textstyle{1\over2}} \left[
       1 + \tanh\left(\beta-\betat \over \Delta\beta\right) \right]
\end {equation}
and determine $\betat$ and $\Delta\beta$ by fitting to our Monte Carlo
results.  We haven't attempted to model the dependence of $\eps$ on
transverse size $T$.  However, for fixed transverse size the model
(\ref{eq:Pfinal}) predicts
\begin {equation}
   \Delta\beta \propto 1/L \,.
\end {equation}

As mentioned in sec.~\ref{sec:method}, there are systematic errors which make
the center of the $\tanh$ curves shift as $L$ is increased.  One example would
occur if the two halves aren't adequately equilibrated at the inverse
temperature $\beta$ before allowing the domain walls to evolve.  Then, even
exactly at $\betat$, there might be some systematic bias to the initial motion
of the walls until thermal equilibrium is reached.  This could be modeled by
simply starting $z$ with some systematic initial offset, {\it i.e.} $z =
L/2+a$.  Secondly, even if proper equilibrium is reached,
the ``stickiness'' of the
domain walls need not be the same in the two phases.  For example, we have
only been discussing the average longitudinal coordinates $z_1$ and $z_2$ of
the domain walls.  In fact, the domain walls have transverse excitations and
might have slightly longer, thinner fingers reaching out into one phase than
into the other.  The separation at which they first touch each other might be
larger in one phase than in the other.  Such an effect could be modeled by
considering the effective end-points of the game to be slightly and
asymmetrically different from $z{=}0$ and $z{=}L$, now being $z{=}b_1$ and
$z{=}L-b_2$.  By a shift of coordinate, this can again be considered as the
problem of starting slightly away from $z=L/2$.

So a model for the systematic error is to ask what happens if the
probability is really $P(L/2+a)$ from (\ref{eq:Pgeneral}), for some
$a$, but we nonetheless tried to extract a value of $\betat$
fitting the form $P(L/2)$ for (\ref{eq:Pfinal}) to the result.
One easily finds that the systematic error in $\betat$ then scales
as $1/L^2$ for large $L$ (assuming fixed transverse dimension and hence a fixed
proportionality between $\eps$ and $\beta-\betat$).


\section{BCC lattices}
\label{app:BCC}

We have made a few simulations on BCC lattices, but not enough to
extract any $x{\to}0$ limits.
Our results are presented in table~\ref{tab:data_bcc}.
The measurements of $C_\pm$ were made on $40^3$ lattices;
$\betat$ was measured on lattices as big as $160\times20^2$ for
$x{=}1.0$ and $80\times40^2$ for $x{=}0.6$.
When we refer to an $L_1\times L_2\times L_3$ BCC lattice,
we mean one with $L_1 L_2 L_3$ unit cells and so $N=2L_1L_2L_3$ sites.
All of our lattices are helical.
Defining helical boundary conditions on a BCC lattice is perhaps
non-standard, and we explain it below.

\begin{table}
\def\cen#1{\multicolumn{1}{c}{$#1$}}
\def\cenb#1{\multicolumn{1}{c|}{$#1$}}
\begin {center}
\tabcolsep=6pt

\begin {tabular}{|llrrr|}                                \hline
$x$   &\cen{\betat}  & \cen{\epsilon_+}
                                 &\cen{C_+} &$C_+/C_-$ \\
      &              & \cen{\epsilon_-}
                                 &\cen{C_-}&           \\ \hline
\multicolumn{5}{|l|}{{\bf BCC:}}                       \\
1     &  0.113752(5) & -2.344(4) &  1.34(24)& 0.112(3) \\
      &              &-5.808(5)& 11.9(3)  &          \\
0.6   &  0.130291(6) & -2.1975(8)&  1.88(4) & 0.104(4) \\
      &              & -4.392(7) & 18.1(5)  &          \\ \hline
\end {tabular}
\end {center}
\caption
    {%
    \label {tab:data_bcc}
    Same as table~\protect\ref{tab:data} but for BCC lattices.
    Densities are given in units of per lattice point, which
    for BCC differs from per unit cell.
    }%
\end{table}


\subsection*{Helical boundary conditions for BCC lattices}
\label{app:helical}

\begin {figure}
\vbox
    {%
    \vspace* {-17pt}
    \begin {center}
        \leavevmode
        
        \epsfbox [150 300 500 500] {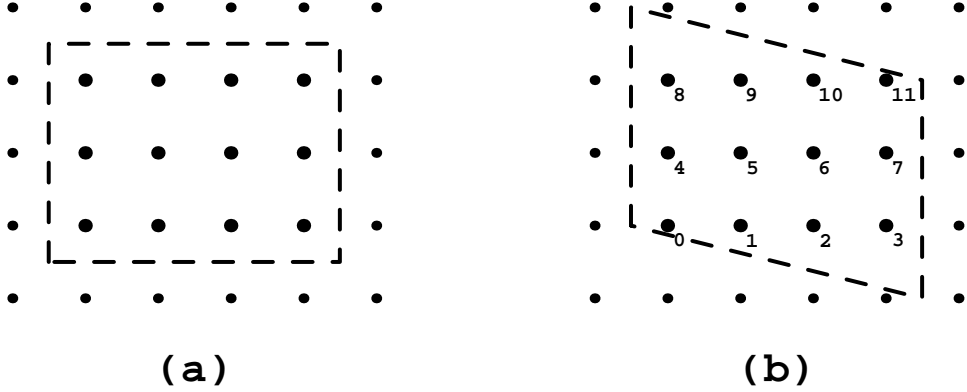}
    \end {center}
    \caption
        {%
        \label {fig:helical-sc}
        Ways to choose a finite volume box from the infinite plane and
        impose periodicity corresponding to (a) normal periodic boundary
        conditions, and (b) helical boundary conditions.
        }%
    }%
\end {figure}

To motivate our definition of helical boundary conditions for BCC lattices,
we first briefly review the situation for simple cubic lattices and, for
simplicity of presentation, will first consider two-dimensional examples.
Fig.~\ref{fig:helical-sc}
shows an infinite simple cubic lattice in two dimensions.
A finite-volume lattice with periodic boundary conditions corresponds to
restricting the system to the sites inside a box, such as shown in
fig.~\ref{fig:helical-sc}a,
and then identifying opposite edges of the box.  The same
volume with helical boundary conditions corresponds instead to the box in
fig.~\ref{fig:helical-sc}b,
again with opposite edges identified.  Helical boundary
conditions retain translation invariance.  Their advantage is that the
sites can be numbered, as shown in fig.~\ref{fig:helical-sc}b,
in such a way that
the offset between one site and its neighbor in a given direction
is always a fixed number modulo $N$, independent of the site chosen,
where $N$ is the total number of sites.  For the two-dimensional lattice
shown, the offsets $\Delta n$ corresponding to the four directions are
\begin {equation}
   \Delta n = \pm 1,\, \pm L_1  ~\mod~ N{=}L_1 L_2 \,.
\end {equation}
In the three-dimensional case, they are
\begin {equation}
   \Delta n = \pm 1,\, \pm L_1,\, \pm L_1 L_2 ~\mod~ N{=}L_1 L_2 L_3 \,.
\label{eq:sc-helical}
\end {equation}
The nature of these offsets is an advantage because, if the lattice is
represented as a linear array in memory, it makes indexing neighbors of
sites quicker and easier than for periodic lattices.
(\ref{eq:sc-helical}) is the definition used in this paper for
an $L_1 \times L_2 \times L_3$ helical simple cubic lattice.

\begin {figure}
\vbox
    {%
    \vspace* {-17pt}
    \begin {center}
        \leavevmode
        
        \epsfbox [150 300 500 500] {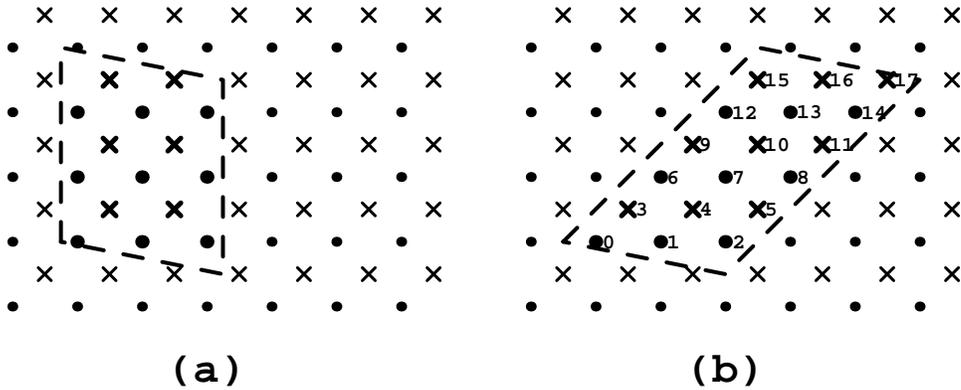}
    \end {center}
    \caption
        {%
        \label {fig:helical-bcc}
        Ways to try to choose a finite volume box for a BCC lattice.
        (a) lacks the advantages of the helical simple cubic
        case; (b) is our definition of a helical BCC lattice.
        }%
    }%
\end {figure}

Fig.~\ref{fig:helical-bcc}a
shows a somewhat similar box drawn for a BCC lattice.%
\footnote{ In two dimensions, an infinite BCC lattice is of course equivalent
   to a simple cubic lattice.  We are discussing it just as a visual aid for
   generalizing to three-dimensional BCC lattices.
}
Unfortunately, there is no way to number the sites in the box so that the
offset $\Delta n$ between neighbors is the same for all sites.  However, if
one instead draws the box as in fig.~\ref{fig:helical-bcc}b, such
numberings are possible.  For two-dimensions, the numbering of
fig.~\ref{fig:helical-bcc}b corresponds to
\begin {equation}
   \Delta n = \pm L_1,\, \pm(L_1-1) ~\mod~ N{=}2 L_1 L_2 \,,
\end {equation}
For three dimensions, where there are 8 nearest-neighbor directions,
the generalization is
\begin {equation}
   \Delta n = \pm L_1L_2,\, \pm(L_1L_2-1),\,
              \pm(L_1L_2-L_1),\, \pm(L_1L_2-L_1-1) ~\mod~ N{=}2 L_1 L_2 L_3 \,.
\end {equation}
This is our definition of an $L_1\times L_2\times L_3$ helical BCC lattice.


\section {A model of ratios and cross-over exponents for small $\alpha$}
\label {app:crossover}

\subsection {Overall scaling}

In this appendix, we elaborate on the assertions of sec.~\ref{sec:crossover}
about how small $x$ needs to be if the Ising critical exponent $\alpha$ is
formally considered small.  We begin by reviewing the case of non-small
$\alpha$ in slightly more detail.  For the sake of specificity, we focus
on the behavior of the correlation lengths $\xi_\pm$.  The correlation
functions $G_\pm(R) = \langle s(R) s(0) \rangle_\pm$ will have the
following form of universal cross-over scaling function near the
transition for small $x$:
\begin {equation}
   G_\pm(R) = b^{-y} G_\pm(b^{y_t}t, b^{y_x}x, b^{-1}R) \,,
\label {eq:G-form}
\end {equation}
where $b$ is the arbitrary renormalization distance scale and
\begin {equation}
   y = d-2+\eta \,,
   \qquad
   y_t = 1/\nu \,,
   \qquad
   y_x = \alpha/\nu
\end {equation}
are the anomalous dimensions of $G_\pm$, $t$, and $x$.
We have ignored the effects of irrelevant operators.
$t$ is a scaling field
corresponding to the reduced temperature.  In this language, the
derivation of the relationship between $\xi_\pm$ and $x$ can be made
by first choosing $b=x^{-1/y_x}$ and writing
\begin {equation}
   G_\pm(R) = x^{y/y_x} G_\pm(x^{-y_t/y_x}t, 1, x^{1/y_x} R) \,.
\end {equation}
As we vary $t$, the transition must occur for some definite value $\tau_0$ of
the first-argument on the left-hand side:
\begin {equation}
   G_\pm(R) = x^{y/y_x} G_\pm(\tau_0,1,x^{1/y_x} R) \,.
\label{eq:G-finessed}
\end {equation}
By dimensional analysis, the long-distance behavior of the right-hand
side of (\ref{eq:G-finessed}) must have the form
\begin{equation}
   G_\pm(\tau_0,1,r) \sim e^{-r/A_\pm} \,,
\end {equation}
and so
\begin{mathletters}%
\begin {eqnarray}
   \xi_\pm \sim A_\pm x^{-1/y_x} \,,
\\
   {\xi_+\over\xi_-} \sim {A_+\over A_-} \sim x^0 \,.
\label {eq:G-xi-ratio}
\end {eqnarray}
\end {mathletters}

Now we can discuss the case where $\alpha$ is small or zero.
(\ref{eq:G-form}) was a special case of a more general situation where
$x$ doesn't scale as a power law.  To be more general, replace
\begin {equation}
   b^{y_x} x \to \bar x(b) \,,
\label{eq:RG replace}
\end {equation}
where $\bar x(b)$ is the solution to some renormalization-group equation.
The non-small $\alpha$ case corresponded to
\begin {equation}
   b \, \partial_b \bar x = y_x \bar x + O(\bar x^2)
     = {\alpha\over\nu} \, \bar x + O(\bar x^2) \,,
\label{eq:RG-x}
\end {equation}
The form (\ref{eq:G-form}) corresponds to ignoring the $O(\bar x^2)$
correction for small $x$, giving a simple, power-law solution for $\bar x$.
When $\alpha=0$, on the other hand, the $O(\bar x^2)$ terms in
(\ref{eq:G-form}) become essential:
\begin {equation}
   b \, \partial_b \bar x = c \, \bar x^2 + O(\bar x^3) \,,
\end {equation}
where we assume $c>0$.  ($c \le 0$ would lead to a second-order Ising
phase transition for small $x>0$ and so is excluded if we assume the
phase transition remains first-order.)
Ignoring the $O(\bar x^3)$ terms, this corresponds to our original scaling
function (\ref{eq:G-form}) with
\begin {equation}
   b^{y_x} x \to {1\over x^{-1} - c\ln b} \,.
\end {equation}
For very small $x$, we now want to choose
\begin {equation}
   b \sim e^{1/cx} \,,
\end {equation}
which gives $\xi_\pm \sim e^{1/cx}$ as discussed in sec.~\ref{sec:crossover}.
However, just as in (\ref{eq:G-xi-ratio}), one will still get
\begin {equation}
   {\xi_+\over\xi_-} \sim {A_+\over A_-} \sim x^0 \,.
\end {equation}

Now we're ready to model the case where $\alpha$ is arbitrarily small but
non-zero.  Clearly we do not want to ignore the $O(\bar x^2)$ term in
the renormalization group equation, so we take
\begin {equation}
   b \, \partial_b \bar x
      = {\alpha\over\nu} \, \bar x + c \, \bar x^2 + O(\bar x^3) \,.
\label{eq:improved RG}
\end {equation}
Ignoring higher-order terms, the solution is
\begin {equation}
   \bar x(b) = {b^{\alpha/\nu} x \over
           1 - (b^{\alpha/\nu}-1){\nu c x \over \alpha} } \,,
\end {equation}
which gives
\begin {equation}
   \xi_\pm \sim \left(
       1 + {\alpha \over \nu cx} 
       \over 1 + {\alpha \over \nu c}
   \right)^{\nu/\alpha} \,,
\end {equation}
which interpolates between the previous cases and
has the properties summarized in sec.~\ref{sec:crossover}.
But, just as in the previous cases, again
\begin {equation}
   {\xi_+\over\xi_-} = {A_+\over A_-} \sim x^0 \,.
\end {equation}
So the ratio is insensitive to the crossover between exponential and
power-law dependence of $\xi_\pm$ on $x^{-1}$.

\subsection {Corrections to scaling}

To address corrections to scaling, supplement (\ref{eq:G-form}) by the
most important irrelevant operator, whose coefficient we shall call
$z$:%
\footnote{
   Some readers may be more familiar with thinking of the RG flow in
   the cubic anisotropy model
   \protect\cite{summary,eps,rudnick,misc-cubic,alford},
   which is in the same universality class, and in terms of the $\epsilon$
   expansion.
   The potential energy in that model is of the form
   $u (\phi_1^2{+}\phi_2^2)^2 + v (\phi_1^4 {+} \phi_2^4)$.
   Roughly speaking, $z$ here corresponds to $v$ and $x$ to a
   linear combination of $-u$ and $v$.
}
\begin {equation}
   G_\pm(R) = b^{-y} G_\pm(b^{y_t}t, b^{y_x}x, b^{-1}R, b^{-\omega}z) \,.
\end {equation}
Now choosing $b$ as before gives
\begin {equation}
   G_\pm(R) = x^{y/y_x} G_\pm(x^{-y_t/y_x}t, 1, x^{1/y_x} R,
                              x^{\omega/y_x} z) \,.
\end {equation}
Quantities like $A_\pm$ now depend on $x^{\omega/y_x} z$ but, for
small $x$, can be Taylor expanded:
\begin {equation}
   {\xi_+\over\xi_-} = {A_+\over A_-}
       = O(x^0) \left[1 + O(x^{\omega/y_x})\right] \,.
\end {equation}

For small or zero $\alpha$, we should make the replacement
(\ref{eq:RG replace}) as before, and turn to the second-order RG
equation (\ref{eq:improved RG}).%
\footnote{
  There will in general also be a term $c' \bar x \bar z$ on the
  right-hand size of (\ref{eq:improved RG}) but, because $z$ is
  irrelevant, this term quickly becomes negligible as $b$ is increased
  and does not affect any of our conclusions.
}
One easily finds that correction-to-scaling laws such as
(\ref{eq:other corrections}), which do not explicitly depend on $\alpha$,
remain valid.


\end {document}